\documentclass[preprint,aps,nofootinbib]{revtex4-1}

\usepackage{bm,amsmath,amssymb,amsfonts,slashed,array,graphicx,soul}
\allowdisplaybreaks
\usepackage[vcentermath,noautoscale]{youngtab}
\usepackage{mathrsfs}
\usepackage{xcolor}
\usepackage{hyperref}
\usepackage{cancel}
\usepackage[normalem]{ulem}
\usepackage{slashed}            
\usepackage{bbm}                
\usepackage{tikz}
\usetikzlibrary{arrows,shapes}
\usetikzlibrary{trees}
\usetikzlibrary{matrix,arrows} 				
\usetikzlibrary{positioning}				
\usetikzlibrary{calc,through}				
\usetikzlibrary{decorations.pathreplacing}  
\usetikzlibrary{decorations.pathmorphing}	
\usetikzlibrary{decorations.markings}
\usetikzlibrary{snakes}
\usepackage[normalem]{ulem}
\tikzset{
    vector/.style={decorate, decoration={snake}, draw},
	provector/.style={decorate, decoration={snake,amplitude=2.5pt}, draw},
	antivector/.style={decorate, decoration={snake,amplitude=-2.5pt}, draw},
    fermion/.style={draw, postaction={decorate},
        decoration={markings,mark=at position .55 with {\arrow[draw]{>}}}},
    fermionbar/.style={draw, postaction={decorate},
        decoration={markings,mark=at position .55 with {\arrow[draw=black]{<}}}},
    fermionnoarrow/.style={draw},
    gluon/.style={decorate, draw,decoration={coil,amplitude=4pt, segment length=6pt}, line width=1},
    scalar/.style={dashed,draw, postaction={decorate},
        decoration={markings,mark=at position .55 with {\arrow[draw]{>}}}},
    scalarbar/.style={dashed,draw, postaction={decorate},
        decoration={markings,mark=at position .55 with {\arrow[draw]{<}}}},
    scalarnoarrow/.style={dash pattern = on 6 pt off 3 pt,draw},
    electron/.style={draw, postaction={decorate},
        decoration={markings,mark=at position .55 with {\arrow[draw]{>}}}},
	bigvector/.style={decorate, decoration={snake,amplitude=4pt}, draw},
	vectorscalar/.style={loosely dotted,draw, postaction={decorate}},
}

\newcommand{\Mpl}{M_{\rm pl}}

\newcommand{\tr}{{\rm Tr}}
\def\lsim{\mathrel{\rlap{\lower4pt\hbox{\hskip1pt$\sim$}}
    \raise1pt\hbox{$<$}}}
\def\gsim{\mathrel{\rlap{\lower4pt\hbox{\hskip1pt$\sim$}}
    \raise1pt\hbox{$>$}}}

\renewcommand{\thefootnote}{\fnsymbol{footnote}}

\begin{document}

\count\footins = 1000

\title{Effective Field Theory of Gravity to All Orders}

\author{Maximilian Ruhdorfer, Javi Serra, and Andreas Weiler }
\email[Email: ]{max.ruhdorfer@tum.de}\email{javi.serra@tum.de}\email{andreas.weiler@tum.de}
\affiliation{Physik-Department, Technische Universit\"at M\"unchen, 85748 Garching, Germany}

\date{\today}

\begin{abstract}
We construct the general effective field theory of gravity coupled to the Standard Model of particle physics, which we name GRSMEFT. Our method allows the systematic derivation of a non-redundant set of operators of arbitrary dimension with generic field content and gravity. We explicitly determine the pure gravity EFT up to dimension ten, the EFT of a shift-symmetric scalar coupled to gravity up to dimension eight, and the operator basis for the GRSMEFT up to dimension eight. Extensions to all orders are straightforward. 
\end{abstract}

\preprint{TUM-HEP-1205-19}

\maketitle

\tableofcontents


\renewcommand{\thefootnote}{\arabic{footnote}}

\section{Introduction}

Effective field theory (EFT) lies at the core of our modern understanding of the fundamental interactions in nature. EFTs encode the dynamics of the relevant degrees of freedom at the scales of interest, and enable the systematic exploration of the effects of heavy states via an infinite set of local operators built out of the light fields. Crucially, the higher the operator's dimension, the smaller the departure it introduces from the leading order dynamics, the latter understood, from this point of view, as a standard quantum field theory.
One of the most elegant quantum field theories is Einstein's theory of general relativity (GR). Formulated in 1915, it has survived all experimental tests, both in measurements on Earth and using precision astrophysical observations at various scales. An important question is how one can systematically test departures from GR, or even at a more basic level, what is the set of independent IR departures one could possibly test. Since most of the probes of GR involve macroscopic distances or very low energies, especially if compared to $(4\pi)M_\text{Pl}$, the maximal scale up to which one could envision GR as a good description of gravitational phenomena, it is natural to work in the language of EFT. 
A subtle problem is that of finding a non-redundant operator basis for the EFT, something that is key in order to properly identify the independent directions in the space of all possible UV completions of the EFT, i.e.~the most general set of physically different deformations of the leading dynamics. This issue is non-trivial because, in general, seemingly independent operators can be related by the equations of motion, partial integration and algebraic identities.
This problem however has been recently solved for the EFT of the Standard Model of particle physics (known as SMEFT)~\cite{Henning:2015alf,Henning:2017fpj}, the method relying on Hilbert series and, to a lesser extent, conformal representation theory.

In this work we extend this solution to account for gravitational interactions, the primary extra ingredient being the identification of the Weyl tensor (and its symmetrized and traceless covariant derivatives) as the building block of the EFT. With the method we develop, we obtain a general and non-redundant set of EFT operators of GR coupled to the SM to all orders, which we call GRSMEFT. Such an EFT is of intrinsic value per se, being the true most general parametrization of all the physically distinct low-energy deviations from the established description of all fundamental interactions known to date, i.e.~the SM dimension-4 Lagrangian and the Einstein-Hilbert term. Besides, one should recall that it has been known since long that GR is non-renormalizable, meaning quantum corrections do in fact require higher-dimensional gravitational operators beyond Einstein-Hilbert, also when matter fields are included. This makes an EFT understanding of the SM and GR practically unavoidable. Furthermore, an EFT of gravity coupled to matter is of relevance in the broader context of physics beyond the SM. For instance, higher-dimensional operators are relevant in inflation or in modified gravity theories (in particular scalar-tensor theories). As an application in this regard, we construct the operator basis for the EFT of a shift-symmetric scalar coupled to gravity. Finally, our method can be of potential use in more formal settings, concerning e.g.~renormalization, scattering amplitudes, or as a systematic connection between UV completions of gravity and low-energy physics.

The paper is structured as follows. In section \ref{sec:method}, we first provide a self-contained review of the Hilbert series method for constructing EFT operator bases, and then apply it to the Euler-Heisenberg Lagrangian as an illustrative example. In section~\ref{sec:gravity}, we extend this method to gravity, showing how we can use the Weyl tensor to construct gravitational EFT operators. In section~\ref{sec:applications}, we apply our approach to the pure gravity EFT up to dimension 10 and to the simple case of a shift-symmetric scalar coupled to gravity up to dimension 8. We comment in passing on known results regarding the renormalization, matching to UV completions, and positivity/causality constraints on these EFTs. Finally, in section~\ref{sec:SMGrav}, we develop the EFT of gravity coupled to the SM to all orders, present its explicit form up to dimension 8 (see also appendix~\ref{app:GRSMEFT8}), and discuss some of its interesting features as well as phenomenological applications and future directions of investigation~\cite{workinprogress}. 


\section{Method} \label{sec:method}

Let us first introduce the key facts about the Hilbert series, to then review the main results of~\cite{Henning:2015alf,Henning:2017fpj} on how the Hilbert series can be used as an efficient tool to find irreducible operator bases for EFTs.%
\footnote{For a physics oriented introduction to the Hilbert series technique we recommend~\cite{Lehman:2015via} (see also~\cite{Jenkins:2009dy}) . For a mathematically more rigorous presentation refer to ~\cite{Neusel,Sturmfels}.}
A reader familiar with these concepts may skip this section.


\subsection{Hilbert Series} \label{subsec:HS}

The Hilbert series $\mathcal{H} (q)$ is a generating function that counts the number of independent group invariants that can be built out of a spurion $q$ in a given representation of the group. It is formally defined as a power series in $q$
\begin{equation}
\mathcal{H} (q) = \sum_{r=0}^\infty c_r\, q^r\,,
\end{equation}
where $c_r$ denotes the number of invariants involving $r$ spurions, with $c_0=1$ by definition. By including multiple spurions $q_i$, one can construct the multi-graded Hilbert series, which provides information on the structure of the invariants. In a field theoretical setting, the spurions stand for field operators $\phi_i$ and derivatives $\mathcal{D}$, i.e.~the Hilbert series in general has the form
\begin{equation}
\mathcal{H}(\mathcal{D},\lbrace\phi_i\rbrace ) = \sum_{r_1,\ldots_,r_n, k} c_{r_1,\ldots ,r_n,k}\, \phi_1^{r_1}\cdots \phi_n^{r_n} \mathcal{D}^k\,,
\end{equation}
where $c_{r_1,\ldots ,r_n,k}$ now indicates the number of invariants of order $k$ in derivates and order $r_i$ in $\phi_i$. As an explicit example, consider a complex scalar field $\phi$ charged under a $U(1)$ symmetry. Any invariant in the scalar potential can be written as a polynomial in the monomial $(\phi^* \phi)$, with each power appearing exactly once. In this case it is straightforward to compute the Hilbert series for the scalar potential, which even has a closed form expression if we formally take the spurions to be small, $(\phi^*\phi) < 1$,
\begin{equation}
\mathcal{H}(\phi ,\phi^*) = 1 + (\phi^*\phi )+ (\phi^*\phi)^2 + \ldots = \sum_{r=0}^\infty (\phi^*\phi)^r = \frac{1}{1- \phi^*\phi}\,.
\end{equation}
Obtaining the Hilbert series in this example was simple only because we already knew the form of the invariants. However, when multiple spurions in different representations of a group $G$ are involved, it is no longer straightforward to find all the invariants. This task can be greatly simplified using group characters. The character of a representation $\mathbf{R}$ of a group $G$ is defined as $\chi_{\mathbf{R}}(g) = \tr_{\mathbf{R}}(g)$ with $g\in G$. Group characters of compact Lie groups are orthonormal w.r.t.~the integration over the group's Haar measure, i.e.~$\int d\mu_G (g)\,  \chi_{\mathbf{R}}(g)\,  \chi_{\mathbf{R'}}^*(g) = \delta_{\mathbf{R}\, \mathbf{R'}}$. Therefore, taking all possible tensor products of the spurions, which amounts to multiplying their characters, and projecting them onto the trivial representation yields all the group invariants. For a bosonic spurion $\phi_{\mathbf{R}}$ in the representation $\mathbf{R}$, the generating function for the characters of all the symmetric tensor products is the plethystic exponential (PE)~\cite{Benvenuti:2006qr,Feng:2007ur}
\begin{equation}
\text{PE}[\phi_{\mathbf{R}}\, \chi_{\mathbf{R}}(z)] = \sum_{n=0}^\infty \phi_{\mathbf{R}}^n\,  \chi_{\text{Sym}^n(\mathbf{R})} (z) = \exp\bigg[\sum_{r=1}^\infty \frac{1}{r} \phi_{\mathbf{R}}^r\, \chi_{\mathbf{R}}(z^r) \bigg]\,,
\label{eq:PE}
\end{equation}
where $\text{Sym}^n(\mathbf{R})$ is the symmetric tensor product of $n$ representations $\mathbf{R}$ and $z = \{z_1, \ldots, z_{\rm{rank}(G)}\}$ are the rank$(G)$ variables parameterizing the group. For a short derivation of this formula see appendix~\ref{app:PE}. The fermionic plethystic exponential (PEF)~\cite{Hanany:2014dia} is the counterpart for fermionic spurions, where the antisymmetric tensor product has to be taken,
\begin{equation}
\text{PEF}[\phi_{\mathbf{R}}\, \chi_{\mathbf{R}}(z)] = \sum_{n=0}^\infty \phi_{\mathbf{R}}^n\,  \chi_{\wedge^n(\mathbf{R})} (z) = \exp\bigg[\sum_{r=1}^\infty \frac{(-1)^{r+1}}{r} \phi_{\mathbf{R}}^r\, \chi_{\mathbf{R}}(z^r) \bigg]\,.
\label{eq:PEF}
\end{equation}
In the following our notation will not differentiate between the fermionic and bosonic version of the PE, as it will be clear from the context which one is meant. For more than one spurion we define the PE as $\rm{PE}[\phi_{\mathbf{R}},\ldots ,\varphi_{\mathbf{R'}}] = \rm{PE}[\phi_{\mathbf{R}}]\cdots \rm{PE}[\varphi_{\mathbf{R'}}]$, where from now on we omit the characters for the spurions in the argument of the PE to ease the notation. From the PE one can obtain the Hilbert series by projecting onto the trivial representation $\mathbf{1}$, with character $\chi_{\mathbf{1}} = 1$, and integrating over the group
\begin{equation}
\mathcal{H}(\phi_{\mathbf{R}},\ldots ,\varphi_{\mathbf{R'}}) = \int d\mu_G\, \text{PE}[\phi_{\mathbf{R}},\ldots ,\varphi_{\mathbf{R'}}]\,.
\label{eq:Molien}
\end{equation}
In the literature this is often referred to as the Molien-Weyl formula (see e.g.~\cite{Sturmfels}). 
Let us illustrate how this machinery works by looking at a simple example with a bosonic spurion $\phi_{\mathbf{2}}$ that transforms in the fundamental representation of $SU(2)$ and its complex conjugate $\phi_{\mathbf{\bar{2}}}^\dagger$. $SU(2)$ has rank one and therefore its characters are a function of one complex variable $y$. The characters for the fundamental $\mathbf{2}$ and adjoint $\mathbf{3}$ representations of $SU(2)$ are $\chi_{\mathbf{\bar{2}}}(y)=\chi_{\mathbf{2}}(y) = y + 1/y$ and $\chi_{\mathbf{3}}(y) = y^2 + 1 + 1/y^2$~\cite{Hanany:2008sb}, while the $SU(2)$ Haar measure can be expressed as a contour integral in the complex plane~\cite{Hanany:2008sb}
\begin{equation}
\int d\mu_{SU(2)} (y) = \frac{1}{2\pi i} \oint_{|y|=1} \frac{dy}{y} \big(1-y^2\big) \,.
\end{equation}
Up to $\mathcal{O}(\phi^2)$, the PE for the spurion $\phi$ is given by
\begin{eqnarray}
\text{PE}[\phi_{\mathbf{2}}] = \exp\bigg[\sum_{r=1}^\infty \frac{1}{r} \phi_{\mathbf{2}}^r\, \chi_{\mathbf{2}}(y^r) \bigg] &=& 1 + \chi_{\mathbf{2}}(y)\,  \phi + \frac{1}{2} (\chi_{\mathbf{2}}(y^2) + \chi_{\mathbf{2}}(y)^2)\, \phi^2+ \mathcal{O}(\phi^3) \nonumber \\
&=& 1 + \chi_{\mathbf{2}}(y)\,  \phi + \chi_{\mathbf{3}}(y) \phi^2 + \mathcal{O}(\phi^3)\,,
\label{eq:PEexample}
\end{eqnarray}
where note that we recover the symmetric part of the $SU(2)$ tensor decomposition $\mathbf{2}\otimes\mathbf{2} = \mathbf{1}_A \oplus \mathbf{3}_S$ from the characters. The PE for $\phi_{\mathbf{\bar{2}}}^\dagger$ is obtained from Eq.~(\ref{eq:PEexample}) after the substitution $\phi_{\mathbf{2}}\rightarrow \phi_{\mathbf{\bar{2}}}^\dagger$. Combining these ingredients and using Eq.~(\ref{eq:Molien}), the Hilbert series up to second order in the fields is given by
\begin{eqnarray}
\mathcal{H}(\phi_{\mathbf{2}}, \phi_{\mathbf{\bar{2}}}^\dagger) &=& \int d\mu_{SU(2)}(y)\, \big( 1 + (\phi_{\mathbf{2}} + \phi_{\mathbf{\bar{2}}}^\dagger ) \chi_{\mathbf{2}}(y) + (\phi_{\mathbf{2}}^2 + \phi_{\mathbf{\bar{2}}}^\dagger\,^2 ) \chi_{\mathbf{3}}(y) + (\phi_{\mathbf{2}} \phi_{\mathbf{\bar{2}}}^\dagger )\chi_{\mathbf{2}}(y) \chi_{\mathbf{2}}(y) + \ldots \big) \nonumber \\
&=& 1 + \phi_{\mathbf{2}} \phi_{\mathbf{\bar{2}}}^\dagger + \mathcal{O} (\phi_{\mathbf{2}}, \phi_{\mathbf{\bar{2}}}^\dagger )^3\,,
\end{eqnarray}
where only the $\phi_{\mathbf{2}} \phi_{\mathbf{\bar{2}}}^\dagger$ term survives the integration, since the tensor product contains one singlet as can be seen from $\chi_{\mathbf{2}}(y) \chi_{\mathbf{2}}(y) = \chi_{\mathbf{1}}(y) + \chi_{\mathbf{3}}(y)$. This result tells us that there is no invariant at the first order in the fields, and exactly one at the second order. This may seem trivial, however by continuing the expansion of the PE to higher orders one can derive the multiplicity and structure of each invariant order by order.


\subsection{Hilbert Series for EFTs} \label{subsec:HSEFT}

The main principle for constructing EFTs is to include all Lorentz and gauge invariant local operators built out of the degrees of freedom accessible at the relevant energy scale. However, to find an operator basis $\mathcal{K} = \lbrace \mathcal{O}_i \rbrace$, i.e.~the minimal set of operators that lead to physically distinct phenomena, is considerably more difficult than just finding all invariants, since in general redundancies appear among operators, which need to be taken care of. Such redundancies appear in two ways: (1) operators proportional to the free field equation of motion (EOM), which can be removed by a field redefinition that leaves the $S$-matrix invariant (see e.g.~\cite{Georgi:1991ch,Arzt:1993gz}), and (2) operators related by a total derivative, which can be transformed into each other using integration by parts (IBP).
The basic building blocks for local operators are fields and derivatives acting on them. For example, for a single scalar field $\phi$, any local EFT operator can be written as a polynomial in $\mathbb{C}[\phi , \partial_\mu\phi ,\partial_\mu\partial_\nu\phi ,\ldots ]$. Monomials such as $\partial_\mu\partial_\nu\phi$ have to be understood as the tensor product of two derivatives acting on the field and therefore still contain a term which is proportional to the free EOM $\partial^2\phi = -m^2 \phi$. These redundant terms ($\phi$ is already a building block) can be avoided by taking only the symmetrized, traceless combination of the derivatives, which we denote as $\partial_{\lbrace\mu_1}\cdots\partial_{\mu_n\rbrace}$.%
\footnote{Note that this remains true even if we replace the derivatives by covariant derivatives. Antisymmetric combinations of covariant derivatives are related to the gauge field strength via $[D_\mu ,D_\nu ] \sim F_{\mu\nu}$. Therefore, the antisymmetric contributions are already accounted for when constructing operators with $F_{\mu\nu}$ and $\phi$.} 
This leads to the single particle module $R_\phi$ as the basic building block~\cite{Henning:2017fpj}
\begin{equation}
R_\phi = \begin{pmatrix}
\phi \\
\partial_\mu \phi\\
\partial_{\lbrace\mu_1}\partial_{\mu_2\rbrace}\phi\\
\vdots
\end{pmatrix}\,.
\label{eq:ScalarSPM}
\end{equation}
One could now use the Molien-Weyl formula with each component of the single particle module as an independent spurion. Using their group characters for the Lorentz representations and integrating over the Lorentz group, one could project out all scalar operators.%
\footnote{The Lorentz group is not a compact Lie group and therefore its characters are not orthonormal. However, since we are not interested in dynamics but only want to enumerate the operators, we can work in Euclidean space, where the Lorentz group $SO(4)\simeq [SU(2)_L\otimes SU(2)_R] / Z_2$ is compact. In addition, since we will be considering fermions, we in fact work with the covering group $Spin(4)$.} 
This would yield an operator basis with the EOM redundancy removed, but the IBP redundancy still present. A procedure which additionally takes care of the IBP redundancy was first proposed in~\cite{Henning:2017fpj}, their main insight the realization that the single particle modules coincide with unitary conformal representations of free fields. The conformal group in four dimensions is isomorphic to $SO(4,2)\simeq SO(6,\mathbb{C})$ and its representations consist of a primary operator $\mathcal{O}_l$ and an infinite tower of derivatives acting on it, its descendants. Schematically, they are of the form 
\begin{equation}
R_{[\Delta ; l]} \sim \begin{pmatrix}
\mathcal{O}_l\\
\partial \mathcal{O}_l \\
\partial^2\mathcal{O}_l\\
\vdots
\end{pmatrix}\,.
\end{equation}
The representations are labeled by the scaling dimension $\Delta$ and the Lorentz representation $l=(l_1,l_2)\in SU(2)_L\times SU(2)_R$ of the primary operator, where $l_i$ denotes the $2\, l_i + 1$ dimensional representation. For a conformal representation to be unitary its scaling dimension $\Delta$ has to satisfy a lower bound $\Delta_l$~\cite{Mack:1975je}
\begin{equation}
\begin{split}
&\Delta \geq \Delta_l = l_1 + l_2 + 2  \quad\qquad l_1\neq 0,\quad l_2\neq 0\,,\\
&\Delta \geq \Delta_l = l_1 + l_2 + 1\quad\qquad l_1 l_2=0\,.
\end{split}
\label{eq:unitarityBound}
\end{equation}
Conformal representations of free fields saturate the unitarity bound, i.e.~$\Delta = \Delta_l$~\cite{Mack:1975je,Barabanschikov:2005ri,Grinstein:2008qk}, which causes some of its descendants to be absent (avoiding negative-norm descendants). Such descendants are exactly those that vanish due to the free EOM. 
This implies that any local operator can now be constructed by taking tensor products of single particle modules, i.e.~tensor products of unitary conformal representations. These tensor products can in turn be decomposed into irreducible conformal representations $\mathcal{O}'$
\begin{equation}
\begin{pmatrix}
\mathcal{O}_l\\
\partial\mathcal{O}_l\\
\partial^2\mathcal{O}_l\\
\vdots
\end{pmatrix}^{\otimes n} = \quad\sum_{\mathcal{O}'} \begin{pmatrix}
\mathcal{O}'\\
\partial\mathcal{O}'\\
\partial^2\mathcal{O}'\\
\vdots
\end{pmatrix}\,.
\end{equation}
The set of all scalar primaries in the tensor product are independent operators with both the IBP and EOM redundancy removed. Therefore, in order to obtain a basis of operators for the EFT, one only has to consider all possible tensor products and project out the scalar $[\Delta,(0,0)]$ representations for all $\Delta$. The corresponding primaries form the EFT basis. Using conformal group characters $\chi_{[\Delta;l]}$, the Hilbert series is schematically 
\begin{equation}
\mathcal{H}\sim \int d\mu_{\rm conformal} \sum_\Delta \chi_{[\Delta;(0,0)]} \, \text{PE}[\lbrace\phi_a\rbrace ]\,.
\end{equation}
Including the integral over possible gauge groups to project out the gauge invariant operators and performing the integral associated with the dilatations one obtains the expression for the Hilbert series%
\footnote{Note that Eq. (\ref{eq:HS}) still holds even if the single particle module is not a unitary conformal representation~\cite{Henning:2017fpj}. However, a closed form expression for $\Delta\mathcal{H}(\mathcal{D},\lbrace\phi_i\rbrace )$ exists only for unitary conformal representations.} (see~\cite{Henning:2017fpj} for details)
\begin{equation}
\mathcal{H}(\mathcal{D},\lbrace\phi_i\rbrace ) = \mathcal{H}_0(\mathcal{D},\lbrace\phi_i\rbrace ) +\Delta \mathcal{H}(\mathcal{D},\lbrace\phi_i\rbrace ) \,,
\label{eq:fullHS}
\end{equation}
with $\mathcal{H}_0(\mathcal{D},\lbrace\phi_i\rbrace )$ given by
\begin{equation}
\mathcal{H}_0(\mathcal{D},\lbrace\phi_i\rbrace ) = \int d\mu_{\text{Lorentz}}(x) \int d\mu_{\text{gauge}}(y) \frac{1}{P(\mathcal{D},x)} \prod_i \text{PE}\bigg[ \frac{\phi_{i}}{\mathcal{D}^{\Delta_{i}}} \bigg]\,,
\label{eq:HS}
\end{equation}
where we denoted the single particle modules by their primaries (i.e.~$\phi_i$ for $R_{\phi_i}$), and recall that $\phi_i$ comes with its character $\chi_{\phi_i}$ in the PE. The group characters for the single particle modules are a product of the conformal and gauge group characters
\begin{equation}
\chi_{\phi_i}(\mathcal{D};x,y) = \chi_{[\Delta_{\phi_i}; l_i]} (\mathcal{D};x)\cdot \chi_{\text{gauge}}(y)\,.
\end{equation}
Furthermore, $\Delta\mathcal{H}(\mathcal{D},\lbrace\phi_i\rbrace )$ in Eq.~(\ref{eq:fullHS}) contains terms of at most scaling dimension 4, and arises from subtleties regarding the orthonormality of the group characters of conformal representations saturating the unitarity bound; basically, the absence of descendants of the form $\Box \mathcal{O}$ and/or $\partial_\mu \mathcal{O}^\mu$ (associated with the EOMs). An explicit expression can be found in~\cite{Henning:2017fpj}. The $1/P(\mathcal{D},x)$ factor corrects for the IBP redundancy, with $P(\mathcal{D},x)$ being the momentum generating function that encodes the information about the symmetric tensor products of derivatives ($\mathcal{D}$ transforming in the fundamental $(\tfrac{1}{2},\tfrac{1}{2})$ representation of the Lorentz group) and is given by (see appendix~\ref{app:PE} for symmetric tensor products)
\begin{equation}
P(\mathcal{D},x) = \sum_{d=0}^\infty \mathcal{D}^d\, \chi_{\text{Sym}^d(1/2,1/2)}(x) = \frac{1}{\text{det}_{(1/2,1/2)}(1-\mathcal{D}\, g)}\,.
\end{equation}
The conformal characters are obtained by tracing over the sum of Lorentz representations in the single particle module weighted with the corresponding scaling dimensions. For instance, for a scalar field with primary scaling dimension $\Delta_\phi=1$ for the primary and single particle module given in Eq.~(\ref{eq:ScalarSPM}), the conformal character is
\begin{equation}
\chi_{[1; (0,0)]} (\mathcal{D};x) = \mathcal{D} (1 - \mathcal{D}^2) \sum_{d=0}^\infty \mathcal{D}^d\, \chi_{\text{Sym}^d(1/2,1/2)}(x) = \mathcal{D}\, P(\mathcal{D},x)  (1 - \mathcal{D}^2)\,.
\label{eq:confCharSc}
\end{equation}
The $\mathcal{D}^1$ factor in Eq.~(\ref{eq:confCharSc}) is due to the scaling dimension of the primary, while each additional power of $\mathcal{D}$ corresponds to a derivative (the subtraction of $\mathcal{D}^2$ in the parenthesis is due to $\Delta_\phi = \Delta_0$ saturating the unitarity bound). Therefore, if each spurion $\phi_i$ in Eq.~(\ref{eq:HS}) is weighted by $\mathcal{D}^{-\Delta_{\phi_i}}$, any occurrence of $\mathcal{D}$ in the Hilbert series will be associated with a derivative. Let us finally note that generically the Hilbert series cannot be computed in full but only as an expansion following a given grading. A common grading is to use the mass dimension $[\phi_i]$ of the operators, i.e.~we rescale the spurions $\phi_i \rightarrow  \epsilon^{[\phi_i]}\phi_i\,, \, \mathcal{D}\rightarrow \epsilon \mathcal{D}$ and expand the Hilbert series in powers of $\epsilon$
\begin{equation}
\mathcal{H}(\mathcal{D} ,\lbrace \phi_i\rbrace ; \epsilon) = \sum_n \epsilon^n\, \mathcal{H}_n (\mathcal{D},\lbrace \phi_i\rbrace )\,.
\end{equation}
Explicit expressions for $P(\mathcal{D},x)$ and for the conformal and gauge characters and the integration measures relevant for this work can be found in appendix~\ref{app:ConfChar}.

We wish to note at this point that the Hilbert series systematically counts the operators at a given order in fields and derivatives, yet it does not explicitly construct them. While knowing the number of operators is exceedingly useful for the latter task -- in fact in this paper this will be information enough to construct the operator basis -- algorithms to directly construct the operators are being developed in the context of the S-matrix \cite{Henning:2019enq,Henning:2019mcv}.


\subsection{Example: Generalized Euler-Heisenberg Lagrangian}\label{subsec:ExEHLagr}

Before moving on to gravity, we will apply this formalism to an instructive example, the generalization of the well-known Euler-Heisenberg Lagrangian, i.e.~we construct the most general EFT for an abelian gauge field.%
\footnote{The original Euler-Heisenberg Lagrangian~\cite{Heisenberg:1935qt} is the EFT for QED at energies much below the electron mass. The CP symmetry of QED forbids CP breaking terms in the Euler-Heisenberg Lagrangian; here we extend it by including CP violating operators.}
The basic building block is the gauge invariant abelian field strength $F_{\mu\nu}$, which satisfies the free EOM 
\begin{equation}
\partial_\mu F^{\mu\nu} = 0\,.
\label{eq:EOMFS}
\end{equation}
From the Bianchi identity $\partial_{[\alpha}F_{\mu\nu ]} = 0$ it also follows that
\begin{equation}
\partial^2 F_{\mu\nu} = 0\,.
\end{equation}
Therefore, the single particle module contains only symmetric and traceless combinations of derivatives of the field strength tensor~\cite{Henning:2017fpj}
\begin{equation}
R_F = \begin{pmatrix}
F_{\mu\nu}\\
\partial_{\lbrace \mu_1} F_{\mu\rbrace \nu}\\
\partial_{\lbrace \mu_1}\partial_{\mu_2} F_{\mu\rbrace \nu}\\
\vdots
\end{pmatrix}\,.
\label{eq:FSSPM}
\end{equation}
The field strength transforms in the reducible $(1,0)\oplus (0,1)$ representation of the Lorentz group. We will therefore work with the combinations $F_{\mu\nu}^{L,R} = \tfrac{1}{2} (F_{\mu\nu} \pm i \tilde{F}_{\mu\nu})$ of the field strength and its dual $\tilde{F}_{\mu\nu} = \tfrac{1}{2} \epsilon_{\mu\nu\rho\sigma} F^{\rho\sigma}$ which live in the $(1,0)$ and $(0,1)$ representations, respectively. The conformal character associated with $F_{\mu\nu}^{L,R}$ is the sum of the characters for the Lorentz representations of the elements in the single particle module in Eq.~(\ref{eq:FSSPM}), weighted by the scaling dimension ($\Delta_{F_{L,R}} = 2$), i.e.~for $F^L_{\mu\nu}$
\begin{equation}
\chi_{[2; (1,0)]} (\mathcal{D};x) = \mathcal{D}^2\, P(\mathcal{D},x)  \big(\chi_{(1,0)}(x) - \chi_{(1/2,1/2)}(x)\, \mathcal{D} + \mathcal{D}^2\big)\,,
\label{eq:confCharFS}
\end{equation}
and the same with $\chi_{(1,0)}(x)$ replaced by $\chi_{(0,1)}(x)$ for $F^R_{\mu\nu}$. The first term in the parenthesis is the Lorentz representation of the conformal primary, i.e.~the field strength, with a tower of symmetrized derivatives generated by $P(\mathcal{D},x)$. The second term subtracts all the descendants where one derivative is contracted with the field strength, corresponding to the Lorentz representation $\partial_\mu F^{L,\, \mu\nu}\sim (\tfrac{1}{2},\tfrac{1}{2})\otimes_A (1,0) = (\tfrac{1}{2},\tfrac{1}{2})$. However, this means that also the term $\partial_\mu \partial_\nu F^{L,\, \mu\nu} \sim (0,0)$ and derivatives thereof are being subtracted, even though they vanish due to the antisymmetry of the field strength and thus were never there from the beginning. For this reason they are added back in the form of the third term in the parenthesis. The structure of Eq.~(\ref{eq:confCharFS}) can also be understood directly in terms of conformal representations~\cite{Barabanschikov:2005ri}. Since abelian field strengths are gauge invariant, the full group characters are $\chi_{F_L}=  \chi_{[2; (1,0)]} (\mathcal{D};x)$ and $\chi_{F_R}=  \chi_{[2; (0,1)]} (\mathcal{D};x)$ and the integral over the gauge group is trivial $\int d\mu_{\text{gauge}} = \int d\mu_{U(1)} = 1$. The Hilbert series in the mass dimension grading scheme, i.e.~$F_{L,R}\rightarrow \epsilon^2\, F_{L,R}$ and $\mathcal{D}\rightarrow \epsilon\, \mathcal{D}$, is thus given by
\begin{eqnarray}
\label{eq:EHHS}
\mathcal{H}_0(\mathcal{D},F_L, F_R;\epsilon) &=& \int d\mu_{\text{Lorentz}}(x) \frac{1}{P(\mathcal{\epsilon\, D},x)} \text{PE}\bigg[\frac{F_L}{\mathcal{D}^{2}},\frac{F_R}{\mathcal{D}^{2}}\bigg] \\
&=& \epsilon^8 \big( F_L^4 + F_L^2 F_R^2 + F_R^4\big) + \epsilon^{10} \big( F_L^4 + F_L^3 F_R + F_L^2 F_R^2 + F_L F_R^3 + F_R^4\big)\mathcal{D}^2 + \ldots\,. \nonumber
\end{eqnarray}
Eq.~(\ref{eq:EHHS}) gives the structure and multiplicity of the operator basis at mass dimension 5 and higher, but it does not reveal how the Lorentz (and gauge) indices of the field strengths and derivatives are contracted. However, if the field content of an operator is known, it is usually straightforward to build Lorentz and gauge invariants. This is especially true if the multiplicity of a given structure is one, since then any non-vanishing contraction can be used as a basis element. The operator basis implied by the Hilbert series in Eq.~(\ref{eq:EHHS}) can be expressed in terms of $F_{\mu\nu}$ and $\tilde{F}_{\mu\nu}$. At mass dimension 8 this is, explicitly, 
\begin{equation}
\mathcal{L} = \frac{c_1}{\Lambda^4} (F_{\mu\nu} F^{\mu\nu})^2 + \frac{c_2}{\Lambda^4} (F_{\mu\nu} \tilde{F}^{\mu\nu})^2 +\frac{i c_3}{\Lambda^4} (F_{\mu\nu} F^{\mu\nu})(F_{\rho\sigma} \tilde{F}^{\rho\sigma}) + \ldots\,,
\end{equation}
where the first two terms also appear in the Euler-Heisenberg Lagrangian. The operator proportional to $c_3$ is CP violating and therefore constitutes an extension of the Euler-Heisenberg Lagrangian. Finally, we note that this method automatically takes algebraic identities, such as $F_{\mu \nu} = - F_{\nu \mu}$ in this simple case, into account. This is because the Hilbert series directly uses group representations to build the invariants, instead of explicitly contracting indices.


\section{Gravity} \label{sec:gravity}

In this section we introduce the Einstein-Hilbert action of GR as the leading contribution at low energies of the EFT of gravity~\cite{Donoghue:1994dn,Burgess:2003jk,Donoghue:2017pgk}. We then identify the relevant building block to construct higher-dimensional, IR subleading operators of the EFT and show that the methods outlined in section~\ref{sec:method} can be applied. In the following we adopt the metric and curvature conventions of~\cite{Donoghue:2017pgk, Gasperini}.


\subsection{General Relativity as an EFT}\label{subsec:GRasEFT}

GR as a classical theory provides an excellent description of gravitational phenomena at large distances. However, once the Einstein-Hilbert action%
\footnote{In the following discussion we will always implicitly assume that the cosmological constant is set to zero.} is quantized
\begin{equation}
S_{\text{EH}} = -\frac{\Mpl^2}{2} \int d^4x \sqrt{-g} R\,,
\label{eq:EHaction}
\end{equation}
where $\Mpl = (8\pi G)^{-1/2}$ is the reduced Planck mass and $R$ the Ricci scalar, it becomes clear that this can only be the leading term in a low-energy EFT. GR as a quantum field theory is non-renormalizable and quantum corrections induce higher-dimensional operators with higher powers of the Riemann tensor~\cite{tHooft:1973bhk,tHooft:1974toh,Goroff:1985th} (the same conclusion is reached when quantum effects from matter fields are considered~\cite{tHooft:1973bhk,tHooft:1974toh,Deser:1974cz,Deser:1974xq,Deser:1974cy}). According to the EFT paradigm, all operators invariant under general coordinate transformations, the gauge symmetry of GR, should be included in a systematic expansion in derivatives over a cutoff scale $\Lambda$, i.e. 
\begin{equation}
S_{eff} =  \int d^4x \sqrt{-g} \bigg[ -\frac{M_{Pl}^2}{2} R + a\, R^2 + b\, R_{\mu\nu} R^{\mu\nu} + c\, R_{\mu\nu\rho\sigma} R^{\mu\nu\rho\sigma} + d\, \Box R + \frac{e}{\Lambda^2} \text{Riem}^3 + \ldots \bigg]\,,
\label{eq:EffGravAction}
\end{equation}
where $\text{Riem}^3$ stands for terms with three Riemann tensors and $\Box = \nabla_\mu \nabla^\mu$ is the contraction of two covariant derivatives. As discussed in the previous section, not all invariants one can write are independent. Operators proportional to the free EOM can be removed by means of field redefinitions. The EOM for GR are the Einstein equations, which can be written in their trace-reversed form as
\begin{equation}
R_{\mu\nu} = \frac{1}{\Mpl^2} (T_{\mu\nu} - \frac{1}{2} T g_{\mu\nu})\,,
\end{equation}
where we included a possible contribution from matter fields through the energy momentum tensor $T^{\mu\nu} = \tfrac{-2}{\sqrt{-g}} \tfrac{\delta S_{\rm matter}}{\delta g_{\mu\nu}}$, with its trace $T= g^{\mu \nu} T_{\mu \nu}$. The free EOM, i.e.~the Einstein equations in vacuum ($T_{\mu\nu}=0$), have the simple solution 
\begin{equation}
R_{\mu\nu} = 0\,.
\end{equation}
This implies that any higher-dimensional operator containing $R_{\mu\nu}$ or $R= g^{\mu\nu} R_{\mu\nu}$ can be eliminated by performing a perturbative field redefinition of the metric 
\begin{equation}
g_{\mu\nu} \rightarrow g_{\mu\nu} + \frac{2}{\Mpl^2} \delta g_{\mu\nu}\,,
\label{eq:MetricRedef}
\end{equation}
which modifies the Einstein-Hilbert action by
\begin{equation}
\delta S_{\rm EH} = \int d^4 x \sqrt{-g} \big[ R^{\mu\nu} - \frac{1}{2} R g^{\mu\nu}\big] \delta g_{\mu\nu} = \int d^4 x \sqrt{-g} R^{\mu\nu} \delta \bar{g}_{\mu\nu}\,,
\label{eq:EHVariation}
\end{equation}
where we introduced the trace-reversed metric perturbation
\begin{equation}
\delta \bar{g}_{\mu\nu} = \delta g_{\mu\nu} - \frac{1}{2} g_{\mu\nu}\, \delta g\,,
\end{equation}
with $\delta g = g^{\mu\nu} \delta g_{\mu\nu}$. From Eq.~(\ref{eq:EHVariation}) it is clear that by choosing an appropriate $\delta \bar{g}_{\mu\nu}$, any operator including $R_{\mu\nu}$ can indeed be removed. Coming back to the effective action in Eq.~(\ref{eq:EffGravAction}), several redundant operators can now be identified (see also~\cite{Burgess:2003jk}). Furthermore, we can drop $\sqrt{-g}\, \Box R = \partial_\mu (\sqrt{-g}\, \nabla^\mu R)$, since it is a total derivative. The term proportional to $R_{\mu\nu\rho\sigma}R^{\mu\nu\rho\sigma}$ can be expressed in terms of $R_{\mu\nu}R^{\mu\nu}$ and $R^2$ because the Gauss-Bonnet term $\mathcal{L}_{GB} = R^2 - 4 R_{\mu\nu} R^{\mu\nu} + R_{\mu\nu\rho\sigma} R^{\mu\nu\rho\sigma}$ is a total derivative in four dimensions; this shifts the Wilson coefficients $a\rightarrow  \tilde{a}= a - c$ and $b\rightarrow  \tilde{b}= b + 4c$. Finally, performing the metric redefinition in Eq.~(\ref{eq:MetricRedef}) with $\delta\bar{g}_{\mu\nu} = -\tilde{b} R_{\mu\nu} - \tilde{a} R g_{\mu\nu}$, gets rid of all the operators with two Riemann structures. The first non-trivial contribution to the gravity EFT appears only at dimension 6 with three Riemann tensors.

In the presence of matter fields, a redefinition of the metric such as Eq.~(\ref{eq:MetricRedef}) also affects the matter action
\begin{equation}
\delta S_{\rm EH} + \delta S_{\rm matter} = \int d^4 x\sqrt{-g} \bigg[ R^{\mu\nu} - \frac{1}{\Mpl^2} ( T^{\mu\nu} - \frac{1}{2} T g^{\mu\nu} ) \bigg] \delta \bar{g}_{\mu\nu}\,.
\end{equation}
We can still use this redefinition to remove any pure gravity terms involving $R_{\mu\nu}$ or $R$, but this will in general introduce mixed curvature-matter operators, such as $R_{\mu\nu} T^{\mu\nu}$. These however usually have a higher mass dimension than the removed operators and can therefore be further removed by an independent redefinition. More care has to be taken when a massive scalar field $\phi$ is involved, since its energy-momentum tensor at leading order in fields and derivatives is $T^{\mu\nu} = \tfrac{1}{2} m^2 \phi^2 g^{\mu\nu} + \ldots$, thus one always introduces new mixed curvature-scalar operators with the same mass dimension as the removed ones. This is in fact not an issue, since the redefinition of the metric can be generalized to include matter fields. To make this clear, consider again the field transformation with $\delta\bar{g}_{\mu\nu} =  -\tilde{b} R_{\mu\nu} - \tilde{a} R g_{\mu\nu}$, which removes the $\tilde{a} R^2 + \tilde{b} R_{\mu\nu} R^{\mu\nu}$ terms in Eq.~(\ref{eq:EffGravAction}), now in the presence of a massive scalar field $\phi$. The change of the action is
\begin{equation}
\delta S_{\rm EH} + \delta S_{\rm matter} = \int d^4 x\sqrt{-g} \bigg[ -\tilde{a}\, R^2 - \tilde{b}\, R_{\mu\nu} R^{\mu\nu} -\frac{m^2}{2\Mpl^2} (\tilde{b} + 4\tilde{a}) \phi^2 R + \ldots \bigg] \,,
\label{eq:varActTot}
\end{equation}
where we dropped higher-dimensional operators. The last term in Eq.~(\ref{eq:varActTot}) is a non-minimal coupling of the scalar field to gravity, which could in fact have been there from the beginning. As anticipated, this term can be removed by a further metric redefinition with $\delta\bar{g}_{\mu\nu}\propto \phi^2 g_{\mu\nu}$. This is a Weyl transformation which takes us to the Einstein frame, where the leading order EOM of the scalar field and gravity are decoupled.

The recipe above lets us, order-by-order in mass dimension, remove any occurrence of $R_{\mu\nu}$ and $R$ in the Lagrangian.%
\footnote{Another comment in this regard is that the same procedure also holds in the presence of classical (gravitational) sources and one is interested in how gravity affects their dynamics; once operators with $R_{\mu\nu}$ and $R$ are removed, one should properly include contact terms between the sources~\cite{Burgess:2003jk}. Besides, we note that in these situations it might be more convenient to work with a non-covariant EFT on the corresponding background metric, as for example in \cite{Goldberger:2004jt,Cheung:2007st,Bloomfield:2012ff,Gleyzes:2013ooa,Franciolini:2018uyq}.}
This implies that the only non-redundant gravitational operators are those built out of the traceless components of the Riemann tensor $R_{\mu\nu\rho\sigma}$. Still, even such operators might not all be independent, due to algebraic identities. The Riemann tensor is cyclic
\begin{equation}
R_{\mu\nu\rho\sigma} + R_{\mu\rho\sigma\nu} + R_{\mu\sigma\nu\rho} = 0
\label{eq:RiemannCycl}
\end{equation}
and it satisfies the Bianchi identity
\begin{equation}
\nabla_\alpha R_{\mu\nu\rho\sigma} + \nabla_\rho R_{\mu\nu\sigma\alpha} + \nabla_\sigma R_{\mu\nu\alpha\rho} = 0\,.
\label{eq:Bianchi}
\end{equation}
Additionally, there are the so-called dimensionally dependent tensor identities, which are obtained by antisymmetrizing tensor indices \cite{Edgar:2001vv}, and can be used to simplify tensor contractions. Once all the redundancies in the gravitational sector are removed, it is clear that field redefinitions of the matter fields can be used to simplify the matter Lagrangian, just as in flat spacetime.

Let us finally briefly comment on spacetimes with torsion. If one is not restricted to a torsion-free spacetime, coupling fermions to gravity will in general induce a non-vanishing torsion tensor $T_{\mu\nu}\,^\rho$. However, even if we chose to include the torsion tensor explicitly as a building block of the EFT, torsion vanishes in vacuum, i.e.~in the free theory, and at the lowest order in derivatives, i.e.~from the leading EOM. Therefore, we conclude that in the presence of matter one can use field redefinitions and work with a torsion-free theory, with shifted coefficients in the matter action \cite{Delacretaz:2014oxa,Carroll:2004st}. In other words, no generality is lost in our EFT by considering a torsion-free spacetime.


\subsection{Building Blocks for the Gravity EFT}\label{subsec:GravityBuildingBlocks}

The Riemann tensor does not transform in an irreducible representation of the Lorentz group. It can be decomposed as
\begin{equation}
R_{\mu\nu\rho\sigma} \sim (1,1)\oplus (2,0) \oplus (0,2) \oplus (0,0)\,,
\end{equation}
where the $(1,1)$ is a symmetric rank-two traceless tensor, identified with the traceless part of the Ricci tensor $R_{\mu\nu}$, the singlet $(0,0)$ is the Ricci scalar $R$, and the component transforming as $(2,0) \oplus (0,2)$ is the Weyl or conformal tensor $C_{\mu\nu\rho\sigma}$. The Weyl tensor is the traceless part of the Riemann tensor and is given by
\begin{equation}
C_{\mu\nu\rho\sigma} \equiv R_{\mu\nu\rho\sigma} - \left( g_{\mu [\rho} R_{\sigma ] \nu}  -g_{\nu [\rho} R_{\sigma ] \mu} \right) +\frac{1}{3} g_{\mu [\rho}g_{\sigma ] \nu} R \,,
\label{eq:WeylTensor}
\end{equation}
where the brackets denote index antisymmetrization, e.g.~$A_{[\mu\nu ]} = \tfrac{1}{2} (A_{\mu\nu} - A_{\nu\mu})$ for arbitrary tensors $A$. It possesses the same symmetries as the Riemann tensor and satisfies the cyclicity and Bianchi identity of Eqs.~(\ref{eq:RiemannCycl}) and (\ref{eq:Bianchi}) up to terms involving $R_{\mu\nu}$ and $R$. As discussed in the previous section, any occurrence of $R_{\mu\nu}$ and $R$ can always be eliminated by an appropriate field redefinition. This leaves the Weyl tensor as the only independent object for constructing gravitational EFT operators. The Einstein equations do not directly constrain the traceless components of the Riemann tensor, but the contracted Bianchi identities imply an EOM for the Weyl tensor, which can be expressed in terms of the Ricci tensor and scalar,
\begin{equation}
\nabla^\mu C_{\mu\nu\rho\sigma} = \nabla_{[\rho} R_{\sigma ] \nu} + \frac{1}{6} g_{\nu [\rho} \nabla_{\sigma ]} R\,.
\end{equation}
For the free theory, i.e.~in vacuum, this simplifies to 
\begin{equation}
\nabla^\mu C_{\mu\nu\rho\sigma}= 0\,,
\end{equation}
in analogy to the EOM for the field strength tensor of gauge fields in Eq.~(\ref{eq:EOMFS}). Additionally, the Bianchi identity in combination with the EOM implies that also $\nabla^2 C_{\mu\nu\rho\sigma}$ is not an independent object, since in vacuum
\begin{equation}
\nabla^2 C_{\mu\nu\rho\sigma} = -2C^\lambda\,_{\mu\rho\alpha} C_{\lambda\nu\sigma}\,^\alpha -2 C^\lambda\,_{\nu\rho\alpha}C_{\mu\lambda\sigma}\,^\alpha - C^\lambda\,_{\alpha\rho\sigma} C_{\mu\nu\lambda}\,^\alpha
\label{eq:boxWeyl}
\end{equation}
plus terms that can be removed due to the EOM. Consequently, the EOM redundancy is taken care of if we consider, similarly to the case of the spin-1 field strength, $C_{\mu\nu\rho\sigma}$ and symmetric traceless combinations of covariant derivatives acting on $C_{\mu\nu\rho\sigma}$ as the basic building blocks for EFT operators. This implies that the single particle module is of the form
\begin{equation}
R_C = \begin{pmatrix}
C_{\mu\nu\rho\sigma}\\
\nabla_{\lbrace \mu_1} C_{\mu\rbrace\nu\rho\sigma}\\
\nabla_{\lbrace \mu_1}\nabla_{\mu_2} C_{\mu\rbrace\nu\rho\sigma}\\
\vdots
\end{pmatrix}\,.
\label{eq:WeylSPM}
\end{equation}
We consider only symmetric combinations of the covariant derivatives, since antisymmetric combinations are related to the Riemann tensor via $[\nabla_\mu ,\nabla_\nu] V^\rho = R_{\mu\nu\sigma}\,^\rho V^\sigma$. An operator containing an antisymmetric combination of covariant derivatives is therefore always equivalent to the tensor product of the Weyl tensor with a descendant that contains fewer derivatives. Analogously to the gauge field strength, we can identify the irreducible representations of the Lorentz group with
\begin{equation}
C_{L/R}^{\mu\nu\rho\sigma} = \frac{1}{2} \bigg( C^{\mu\nu\rho\sigma} \pm i\, \tilde{C}^{\mu\nu\rho\sigma}\bigg)\,,
\label{eq:LRWeyl}
\end{equation}
where the dual Weyl tensor $\tilde{C}^{\mu\nu\rho\sigma} = \epsilon^{\mu\nu\alpha\beta} C_{\alpha\beta}\,^{\rho\sigma}/2$.%
\footnote{The normalization of the Levi-Civita tensor is such that $\epsilon^{0123} = 1/\sqrt{-g}$.} 
$C_{L/R}$ transform in the $(2,0)$ and $(0,2)$ representations of the Lorentz group, respectively. Note that the single particle modules $R_{C_{L/R}}$ cannot be identified with unitary conformal representations in four dimensions. The mass dimension of the Weyl tensor is $[C_{L/R}] = 2$, which violates the unitarity bound for the scaling dimension $\Delta \geq \Delta_{(2,0)} = \Delta_{(0,2)} = 3$ in Eq.~(\ref{eq:unitarityBound}). However, since our aim is only to enumerate and construct operators, we can formally assign a conformal scaling dimension of $\Delta_{C_{L/R}} = 3$ to the spurion representing the Weyl tensor.%
\footnote{Besides, note that Eq.~(\ref{eq:HS}) for the construction of the Hilbert series also holds for single particle modules that are not conformal representations~\cite{Henning:2017fpj}. The advantage of promoting the Weyl tensor to be formally a unitary conformal representation is that in this case there is a closed form expression for $\Delta \mathcal{H}$ in Eq.~(\ref{eq:fullHS}).}
When expanding the Hilbert series we can choose a grading in which the spurion for the Weyl tensor is assigned a weight according to the Weyl tensor's actual mass dimension in four dimensions.
The conformal representations $[3;(2,0)]$ and $[3;(0,2)]$ saturate the unitarity bound and therefore are representations with all descendants proportional to the free EOM $\nabla^\mu C_{\mu\nu\rho\sigma}=0$ (as well as $\nabla^2 C_{\mu\nu\rho\sigma}$) being absent. This is exactly of the form of the single particle module in Eq.~(\ref{eq:WeylSPM}). The corresponding conformal character is
\begin{equation}
\chi_{[3; (2,0)]} (\mathcal{D};x) = \mathcal{D}^3\, P(\mathcal{D},x)  \big(\chi_{(2,0)}(x) - \chi_{(3/2,1/2)}(x)\, \mathcal{D} + \chi_{(1,0)}(x)\, \mathcal{D}^2\big)\,,
\label{eq:confCharWeyl}
\end{equation}
and equivalently for $[3; (0,2)]$. Note that here and in the following the spurion $\mathcal{D}$ denotes covariant derivatives $\nabla_\mu$. 
The structure of Eq.~(\ref{eq:confCharWeyl}) is completely analogous to that of the conformal character for a gauge field strength, Eq.~(\ref{eq:confCharFS}). Using this character, in the next section we will construct the operator basis for EFTs which involve gravity.
Note that similarly to the example in section~\ref{subsec:ExEHLagr}, the Hilbert series method automatically factors in redundancies due to Bianchi identities, cyclicity of indices or dimensionally dependent identities, since we do not construct index contractions but work directly with group representations and form invariants.
In appendix~\ref{app:HigherDim} we generalize Eq.~(\ref{eq:confCharWeyl}) to $d$ spacetime dimensions, pointing out the main difference with respect to the derivation for $d=4$, namely that the single particle module for the Weyl tensor, $R_C$ in Eq.~(\ref{eq:WeylSPM}), cannot be embedded for $d > 4$ in a free field unitary conformal representation. We count and identify as well the basis of effective operators for pure gravity in $d = 5$.


\section{Applications} \label{sec:applications}

In this section we combine the formalism outlined in section~\ref{sec:method} with the considerations of section~\ref{sec:gravity} to construct operator bases for EFTs involving gravity. First we verify and extend the operator basis for gravity in vacuum as given in e.g.~\cite{Endlich:2017tqa}. Next, as a first step towards including matter fields, we build the EFT for a shift-symmetric scalar coupled to gravity and point out redundancies in the operator basis of \cite{Solomon:2017nlh}. Finally, we list for the first time the complete basis of the SM coupled to gravity.


\subsection{Gravity in Vacuum}\label{subsec:gravityVacuum}

In vacuum the only independent operators that do not vanish on-shell are the Weyl tensor and its dual, which can be used to form the chiral combinations $C_{L/R}^{\mu\nu\rho\sigma}$, as shown in Eq.~(\ref{eq:LRWeyl}). The building blocks are therefore their corresponding single particle modules, which can be embedded into conformal representations if we formally assign to $C_{L/R}$ a conformal scaling dimension of $\Delta_{C_{L/R}}=3$. Hence, their group characters are
\begin{equation}
\chi_{C_L} (\mathcal{D}; x)= \chi_{[3,(2,0)]} (\mathcal{D}; x)\,,\qquad \chi_{C_R} (\mathcal{D}; x)= \chi_{[3,(0,2)]} (\mathcal{D}; x)\,,
\end{equation}
with the explicit form of the conformal characters given in Eq.~(\ref{eq:confCharWeyl}) and appendix~\ref{app:ConfChar}. Grading the spurions according to their actual mass dimension, i.e.~$C_{L/R}\rightarrow \epsilon^2\, C_{L/R}$ and $\mathcal{D}\rightarrow \epsilon\, \mathcal{D}$, the Hilbert series can be computed as an expansion in mass dimension using Eq.~(\ref{eq:fullHS})%
\footnote{For unitary conformal representations, as it is the case here, $\Delta \mathcal{H}$ can be evaluated explicitly~\cite{Henning:2017fpj} and yields $-\epsilon^4 \mathcal{D}^4$, which cancels the $+\epsilon^4 \mathcal{D}^4$ that one obtains from evaluating the integral over the group measure.} 
\begin{eqnarray}
\mathcal{H}(\mathcal{D},C_L ,C_R; \epsilon) &=& \int d\mu_{\text{Lorentz}}(x) \frac{1}{P(\mathcal{\epsilon\, D},x)} \text{PE}\bigg[\frac{C_L}{\epsilon\, \mathcal{D}^{3}},\frac{C_R}{\epsilon\, \mathcal{D}^{3}}\bigg] + \Delta \mathcal{H}(\mathcal{D},C_L ,C_R; \epsilon) \nonumber \\
\label{eq:GravityHilbert}
&=&\epsilon^4 \left( C_L^2 + C_R^2\right) + \epsilon^6\left( C_L^3 + C_R^3\right) + \epsilon^8\left( C_L^4 + C_L^2 C_R^2 + C_R^4 \right) \\
&&+ \,\epsilon^{10}\left(C_L^5 + C_L^3 C_R^2 + C_L^2 C_R^3 + C_R^5 + C_L^4 \mathcal{D}^2 + C_L^2 C_R^2 \mathcal{D}^2 + C_R^4 \mathcal{D}^2 \right) +\ldots\,. \nonumber
\end{eqnarray}
The terms at $\mathcal{O}(\epsilon^4)$ correspond to the operators $C_{\mu\nu\rho\sigma} \tilde{C}^{\mu\nu\rho\sigma}$ and $C_{\mu\nu\rho\sigma} C^{\mu\nu\rho\sigma}$. Both can be dropped, since the first is a total derivative and the second can be related to $R_{\mu\nu}R^{\mu\nu}$ and $R^2$ because the Gauss-Bonnet term is a total derivative in four dimensions (see section~\ref{subsec:GRasEFT}). These operators were misidentified as being non-redundant, since they are in fact related to gravitational topological terms. 
That topological terms are misidentified by our method was already realized in \cite{Henning:2017fpj}, being a consequence of working with covariant field strengths instead of gauge fields. The operators $C_{\mu\nu\rho\sigma} \tilde{C}^{\mu\nu\rho\sigma}$ and $C_{\mu\nu\rho\sigma} C^{\mu\nu\rho\sigma}$ give rise, respectively, to the four-dimensional Pontryagin and Euler densities~\cite{Eguchi:1976db}.
The other terms in Eq.~(\ref{eq:GravityHilbert}) indicate the structure and multiplicity of the basis of operators for the most general gravity Lagrangian in vacuum up to mass dimension 10. In terms of the Weyl tensor and its dual the basis can be written as
\begin{eqnarray}
\label{eq:GravityVacuumAction}
S = \int d^4x \sqrt{-g} \bigg[&-&\frac{\Mpl^2}{2} R + \frac{c_1}{\Lambda^2} \mathcal{I} + \frac{c_2}{\Lambda^2} \tilde{\mathcal{I}} + \frac{d_1}{\Lambda^4} \mathcal{C}^2 + \frac{d_2}{\Lambda^4} \mathcal{C}\tilde{\mathcal{C}} + \frac{d_3}{\Lambda^4} \tilde{\mathcal{C}}^2 \\
&+& \frac{e_1}{\Lambda^6} \mathcal{I}\mathcal{C} + \frac{e_2}{\Lambda^6} \tilde{\mathcal{I}}\mathcal{C} + \frac{e_3}{\Lambda^6} \mathcal{I}\tilde{\mathcal{C}} + \frac{e_4}{\Lambda^6} \tilde{\mathcal{I}}\tilde{\mathcal{C}} + \frac{e_5}{\Lambda^6} \mathcal{F}\mathcal{C}+ \frac{e_6}{\Lambda^6} \mathcal{F}\tilde{\mathcal{C}} + \frac{e_7}{\Lambda^6} \tilde{\mathcal{F}}\tilde{\mathcal{C}} + \ldots \bigg]\,, \nonumber
\end{eqnarray}
with the basic invariants
\begin{align}
&\mathcal{I} = C_{\mu\nu}\,^{\rho\sigma} C^{\mu\nu\alpha\beta} C_{\alpha\beta\rho\sigma}\,,\qquad &\tilde{\mathcal{I}} &= C_{\mu\nu}\,^{\rho\sigma} C^{\mu\nu\alpha\beta} \tilde{C}_{\alpha\beta\rho\sigma}\,, \label{eq:gravitySH1}\\
&\mathcal{C} = C_{\mu\nu\rho\sigma} C^{\mu\nu\rho\sigma}\,, &\tilde{\mathcal{C}} &= C_{\mu\nu\rho\sigma} \tilde{C}^{\mu\nu\rho\sigma}\,,
\label{eq:gravitySH2}\\
&\mathcal{F} = (\nabla_\alpha C_{\mu\nu\rho\sigma}) (\nabla^\alpha C^{\mu\nu\rho\sigma})\,, &\tilde{\mathcal{F}} &= (\nabla_\alpha C_{\mu\nu\rho\sigma}) (\nabla^\alpha \tilde{C}^{\mu\nu\rho\sigma})\,. \label{eq:gravitySH3}
\end{align}
The first line of Eq.~(\ref{eq:GravityVacuumAction}) is equivalent to the effective action in Eq.~(1.1) of~\cite{Endlich:2017tqa}.%
\footnote{In \cite{Endlich:2017tqa} the operator basis is given in terms of the Riemann tensor, which after the decomposition in Eq.~(\ref{eq:WeylTensor}) coincides with the operators we found in Eq.~(\ref{eq:GravityVacuumAction}) modulo terms with $R_{\mu\nu}$ and $R$, which can be removed by field redefinitions as explained in section~\ref{subsec:GRasEFT}.}
The second line of Eq.~(\ref{eq:GravityVacuumAction}) shows for the first time the basis of gravitational operators at dimension 10. We used the Invar package \cite{MartinGarcia:2008qz} to explicitly construct and classify the (CP even) operators and to check that they are indeed independent.
We note that in general the basis of operators without derivatives corresponds to the most general polynomial of the invariants $\mathcal{C}$, $\tilde{\mathcal{C}}$, $\mathcal{I}$ and $\tilde{\mathcal{I}}$, these being the four scalar quantities that completely determine the spacetime curvature in four dimensions (see e.g.~the discussion in~\cite{Landau:1987gn}).
This fact is further verified by computing the Hilbert series for the left- and right-handed Weyl tensors $C_{L/R}$ without derivatives, which according to the Molien-Weyl formula Eq.~(\ref{eq:Molien}) yields
\begin{align}
\mathcal{H}(C_L, C_R) = \frac{1}{(1-C_L^2)(1-C_R^2)(1-C_L^3)(1-C_R^3)} \,.
\end{align}
Therefore all operators without derivatives are generated by the four basic invariants $C_L^2$, $C_R^2$, $C_L^3$ and $C_R^3$, corresponding to the invariants in Eqs.~(\ref{eq:gravitySH1}) and~(\ref{eq:gravitySH2}).

It is straightforward to compute even higher order contributions to the Hilbert series with this formalism. Finding the explicit form for the corresponding basis of operators can be more involved. However, since we know how many independent operators appear in each category, one does not need to classify all invariants. It is sufficient to find as many operators as the multiplicity in the Hilbert series predicts and check that they are independent.

Let us finally comment on some important aspect of the gravity EFT in vacuum. At one loop GR is finite~\cite{tHooft:1974toh}, a fact that from the EFT perspective follows from naive dimensional analysis (NDA) and the absence of non-redundant operators at $O(\epsilon^4)$. At two loops the Einstein-Hilbert term induces renormalization group (RG) evolution of the CP preserving cubic curvature term in Eq.~(\ref{eq:GravityVacuumAction}), with running coefficient~\cite{Goroff:1985th,vandeVen:1991gw,Bern:2017puu} 
\begin{align}
\mu \frac{\partial c_1}{\partial \mu} = \frac{1}{120} \frac{\Lambda^2}{\Mpl^2} \frac{1}{ (4\pi)^4}\,.
\end{align} 
Heavy matter fields (scalars, fermions or vectors) contribute at one loop to the gravity EFT~\cite{Avramidi:1986mj,Goon:2016mil}, giving rise to a finite contribution to $c_1$, which for e.g.~a Dirac fermion of mass $\Lambda$ reads $c_1 = -\tfrac{1}{7560} (1/4\pi)^2$, as well as to contributions to several other operators that are not present in our basis being dependent on $R_{\mu \nu}$, $R$.%
\footnote{Matter fields also contribute to the RG running of $\Mpl$ and the cosmological constant.} 
Gravitational UV completions, such as (super-)string theories, generate as well a specific pattern of Wilson coefficients below the string scale, see e.g.~\cite{Gross:1986iv}.
In a generic gravity effective Lagrangian, the Wilson coefficients are arbitrary $\mathcal{O}(1)$ numbers. 
To be more precise, the size of the coefficient can be estimated following NDA as
\begin{equation}
\mathcal{L}_{eff} = \frac{m_*^4}{g_*^2} L \! \left( \frac{R_{\mu\nu\rho\sigma}}{m_*^2}, \frac{\nabla_\mu}{m_*}\right) - \frac{\hat M_{\rm{pl}}^2}{2} R\,,
\label{GRNDA}
\end{equation}
where $m_*$ and $g_*$ broadly characterize, respectively, the typical mass scale and coupling of the UV resonances that have been integrated out. We have explicitly included a ``fundamental'' Einstein-Hilbert term, to distinguish between a bona fide completion of GR such as string theory, which would correspond to $\Mpl \sim \hat M_{\rm{pl}} \sim m_*/g_*$, and the case where the graviton can be considered external, i.e.~``elementary'', to the dynamics giving rise to $L$, e.g.~loops of $N$ matter particles of mass $m_*$, for which $g_* \sim 4 \pi / \sqrt{N}$. The simple power counting of Eq.~(\ref{GRNDA}) implies then $c_1 = \mathcal{O}(1)$ for $\Lambda \sim g_* m_*$, and similarly for the rest of Wilson coefficients.
Interestingly, several works have derived constraints on the sign and size of such coefficients based on causality, unitarity and analyticity. For instance, positivity of the coefficients of the CP even dimension 8 operators, i.e.~$d_1,d_3>0$, has been derived based on causality of graviton propagation~\cite{Gruzinov:2006ie}, or unitarity and analyticity of graviton scattering amplitudes~\cite{Adams:2006sv,Bellazzini:2015cra}, while \cite{Endlich:2017tqa} extended the former causality analysis to the CP odd operator, concluding $d_2 \lesssim d_1 d_3$. We should recall however that these arguments are delicate when applied to gravitational interactions, in particular~\cite{Adams:2006sv,Bellazzini:2015cra} neglect the universal $t$-channel singularity due to graviton exchange, a fact that could be justified by e.g.~the rationale presented in~\cite{Bellazzini:2019xts}.
Besides, \cite{Camanho:2014apa} argues that in a weakly coupled theory of gravity, with $g_* \sim m_*/\Mpl \ll 1$, avoiding causality violation originating from cubic curvature terms ($\mathcal{I}$ and $\tilde{\mathcal{I}}$ in Eq.~(\ref{eq:GravityVacuumAction}), with coefficients $c_1$ and $c_2$ effectively of tree-level size), requires an infinite tower of higher-spin states to appear at or near the EFT cutoff $m_*$, regardless of the coefficients sign. Instead, for other types of UV completions where the graviton is elementary, with $\Mpl \gg m_*/g_*$ such as from loops of matter particles~\cite{Goon:2016mil}, acausality lies beyond the validity of the EFT.


\subsection{Shift-Symmetric Scalar Coupled to Gravity}\label{subsec:ScalarShiftSymmetric}

As a second application of our method we consider a shift-symmetric scalar $\phi$ coupled to gravity. The shift symmetry $\phi \rightarrow \phi +\alpha$ implies that the scalar can only couple derivatively, i.e.~it will always appear with at least one derivative acting on it. The scalar can be thought of as the massless Nambu-Goldstone boson of a spontaneously broken $U(1)$ symmetry. Because of this interpretation, it is no surprise that the single particle module for the shift-symmetric scalar has the same form as for non-linear field realizations as given in \cite{Henning:2017fpj}%
\footnote{\cite{Henning:2017fpj} used the decomposition of the Maurer-Cartan form $U^{-1}\partial_\mu U= u_\mu^i X^i + v_\mu^a T^a = u_\mu + v_\mu$ into components along broken generators $X^i$ and unbroken generators $T^a$ to obtain a linearly transforming building block $u_\mu$ from the non-linearly transforming Goldstone matrix $U=\exp (i \phi^i X^i / f_{\phi})$. For a spontaneously broken $U(1)$ symmetry $u_\mu \propto \partial_\mu\phi$. Here the only difference is in the mass dimension, $[\nabla_\mu \phi] = 2$ whereas $[u_\mu ]=1$.}
\begin{equation}
R_{d\phi} = \begin{pmatrix}
\nabla_\mu \phi\\
\nabla_{\lbrace \mu_1} \nabla_{\mu\rbrace} \phi\\
\nabla_{\lbrace \mu_1} \nabla_{\mu_2} \nabla_{\mu\rbrace} \phi\\
\vdots
\end{pmatrix}\,.
\label{eq:SPMshiftsymmscalar}
\end{equation}
In Eq.~(\ref{eq:SPMshiftsymmscalar}) we already imposed the scalar's EOM $\nabla_\mu \nabla^\mu \phi = 0$. Therefore, the weighted character for the single particle module is identical to the one for non-linear realizations \cite{Henning:2017fpj}
\begin{equation}
\chi_{d\phi} (\mathcal{D}; x) = \mathcal{D} \big[ \big(1 -\mathcal{D}^2\big) P(\mathcal{D}; x) -1\big]\,.
\end{equation}
Note that this single particle module is not a conformal representation since the primary field is a total derivative. However, we can still use Eq.~(\ref{eq:HS}) to construct the Hilbert series for higher-dimensional operators, which are the ones we are interested in. To obtain the full Hilbert series, i.e.~including also operators of dimension 4 and lower, we have to rely on the results of~\cite{Henning:2017fpj} for non-linear realizations.
A basis for the CP even operators, up to 6 derivatives, was constructed in~\cite{Solomon:2017nlh}. We compute the Hilbert series for operators with 6 and 8 derivatives and compare our basis to their results.
The Hilbert series as an expansion in derivatives can be obtained by rescaling the spurions $C_{L/R} \rightarrow \epsilon^2\, C_{L/R}$, $\mathcal{D}\rightarrow \epsilon\, \mathcal{D}$ and $d\phi \rightarrow \epsilon d\phi$ and expanding in $\epsilon$
\begin{equation}
\mathcal{H}_0(\mathcal{D},C_L ,C_R, d\phi ; \epsilon) = \int d\mu_{\text{Lorentz}}(x) \int \frac{1}{P(\mathcal{\epsilon\, D},x)} \text{PE}\bigg[\frac{C_L}{\epsilon\, \mathcal{D}^{3}},\frac{C_R}{\epsilon\, \mathcal{D}^{3}},\frac{d\phi}{\epsilon\, \mathcal{D}^{2}}\bigg] = \sum_n \epsilon^n \mathcal{H}_n \,.
\end{equation}
The 6-derivative Hilbert series is
\begin{equation}
\mathcal{H}_{6} = C_L^3+C_R^3 + d\phi^3 C_L \mathcal{D}+d\phi^3 C_R \mathcal{D}+d\phi^2 C_L^2+d\phi^2 C_R^2+d\phi^6+d\phi^4 \mathcal{D}^2\,.
\end{equation}
If we restrict to CP even operators, the EFT operator basis at the 6-derivative level can be written as
\begin{eqnarray}
&\mathcal{O}_1 = \big[(\nabla_\mu \phi)^2\big]^3\,, \qquad \mathcal{O}_2 = (\nabla_\mu \phi)^2 (\nabla_{\rho}\nabla_{\sigma}\phi)^2\,,\qquad \mathcal{O}_3 = C_{\mu\nu}\,^{\rho\sigma} C^{\mu\nu\alpha\beta} C_{\alpha\beta\rho\sigma}\,,& \nonumber \\
&\mathcal{O}_4 = (C_{\alpha\beta\rho\sigma})^2 (\nabla_\mu\phi )^2\,,\qquad \mathcal{O}_5 = C_{\mu\nu\rho\sigma} (\nabla^\mu\phi)(\nabla^\rho\phi )(\nabla^{\nu}\nabla^{\sigma}\phi)\,.&
\label{eq:6derScalarBasis}
\end{eqnarray}
Note that \cite{Solomon:2017nlh} lists two additional operators in their operator basis (written in terms of the Riemann instead of the Weyl tensor), $(\nabla_\alpha R_{\mu\nu\rho\sigma})^2$ and $R_{\mu\nu\alpha\beta} (\nabla^\mu\nabla^\alpha \phi)(\nabla^\nu\nabla^\beta\phi )$. Both of these operators are redundant: the first is related to $\mathcal{O}_3$, whereas the second to $\mathcal{O}_4$ (see appendix~\ref{app:redundOp}). Therefore, the correct CP even operator basis for the 6-derivative Lagrangian is that in Eq.~(\ref{eq:6derScalarBasis}).
At 8 derivatives we find 26 independent operators, with the structures as given in the Hilbert series
\begin{equation}
\begin{split}
\mathcal{H}_8 =\, & C_L^4+ C_L^2 C_R^2+C_R^4+d\phi^8+2d\phi^6 \mathcal{D}^2+d\phi^5 \mathcal{D}^3+d\phi^4 \mathcal{D}^4+ d\phi^5 \mathcal{D} C_L+d\phi^5 \mathcal{D} C_R\\
&+d\phi^4 C_L C_R+d\phi^4 \mathcal{D}^2 C_L+d\phi^4 C_L^2 +d\phi^4 \mathcal{D}^2 C_R +d\phi^4 C_R^2+2d\phi^3 \mathcal{D} C_L^2\\
&+2d\phi^3 \mathcal{D} C_R^2+d\phi^2 \mathcal{D}^2 C_L C_R+2d\phi^2 
\mathcal{D}^2 C_L^2 +d\phi^2 C_L^3 +2d\phi^2 \mathcal{D}^2 C_R^2+d\phi^2  C_R^3\,.
\end{split}
\end{equation}
Finally, we note that there are two operators that respect the shift-symmetry that cannot be found by our method (and are missing in~\cite{Solomon:2017nlh}):
\begin{equation}
\phi C_{\mu\nu\rho\sigma} C^{\mu\nu\rho\sigma} \,, \qquad \phi C_{\mu\nu\rho\sigma} \tilde{C}^{\mu\nu\rho\sigma}\,.
\label{eq:ScalarTop}
\end{equation}
Not surprisingly, these operators are related to (gravitational) topological terms.


\section{Standard Model Coupled to Gravity}\label{sec:SMGrav}

Let us now construct the complete EFT for the SM, i.e.~all operators including SM fields and gravity. Operators including gravity are usually omitted in the SMEFT, even though gravity is part of the SM. We work with one generation of fermions and introduce them in the left-handed $(\tfrac{1}{2},0)$ representation of the Euclidean Lorentz group $SO(4)\simeq SU(2)_L\times SU(2)_R$ along with their right-handed conjugates.%
\footnote{This means we work with the charge conjugated fields of the standard SM right-handed fermions, i.e.~$u^c, d^c, e^c$. In the following we will drop the superscript $c$.}
In the following we adopt the notation of \cite{Henning:2015alf} and denote the spurion fields as
\begin{equation}
\lbrace \phi_a\rbrace = \left\{H,H^{\dagger },B_L,B_R,W_L,W_R,G_L,G_R,C_L,C_R,Q,Q^{\dagger },u,u^{\dagger },d,d^{\dagger },L,L^{\dagger },e,e^{\dagger }\right\}\,,
\end{equation} 
with their representation under the Lorentz and SM gauge group $SU(3)_C\times SU(2)_W\times U(1)_Y$ given in Table \ref{tab:FieldsRepr}.
\begin{table}[t!]
\centering
\begin{tabular}{ c | *4c }    
~		&	$\quad SU(2)_L\times SU(2)_R\quad$		&	$SU(3)_C$	&	$SU(2)_W$		&	$U(1)_Y$	\\ \hline
$H$		&	$(0,0)$					&	$\mathbf{1}$	&	$\mathbf{2}$		&	$1/2$\\
$B_L$	&	$(1,0)$					&	$\mathbf{1}$	&	$\mathbf{1}$		&	$0$\\
$W_L$	&	$(1,0)$					&	$\mathbf{1}$	&	$\mathbf{3}$		&	$0$\\
$G_L$	&	$(1,0)$					&	$\mathbf{8}$	&	$\mathbf{1}$		&	$0$\\
$C_L$	&	$(2,0)$					&	$\mathbf{1}$	&	$\mathbf{1}$		&	$0$\\
$Q$		&	$(\tfrac{1}{2},0)$			&	$\mathbf{3}$	&	$\mathbf{2}$		&	$1/6$\\
$u_c$	&	$(\tfrac{1}{2},0)$		&	$\mathbf{\bar{3}}$	&	$\mathbf{1}$	&	$-2/3$\\
$d_c$	&	$(\tfrac{1}{2},0)$		&	$\mathbf{\bar{3}}$	&	$\mathbf{1}$	&	$1/3$\\
$L$			&	$(\tfrac{1}{2},0)$		&	$\mathbf{1}$		&	$\mathbf{2}$	&	$-1/2$\\
$e_c$	&	$(\tfrac{1}{2},0)$		&	$\mathbf{1}$		&	$\mathbf{1}$	&	$1$
\end{tabular}
\caption{Representations of the spurions under the Lorentz and SM gauge groups.} 
\label{tab:FieldsRepr}
\end{table}
We can write the Hilbert series as
\begin{equation}
\mathcal{H}(\lbrace \phi_a\rbrace ; \mathcal{D} ) = \int d\mu_\text{gauge} (y) \int d\mu_{\text{Lorentz}}(x)\, \frac{1}{P (\mathcal{D} ;x)} \text{PE}\bigg[\bigg\lbrace \frac{\phi_a}{\mathcal{D}^{\Delta_a}} \bigg\rbrace \bigg]\,,
\end{equation}
with the integral over the gauge groups given by
\begin{equation}
\int d\mu_\text{gauge} (y) = \int d\mu_{U(1)_Y}(v) \int d\mu_{SU(2)_W}(w) \int d\mu_{SU(3)_C}(z_1,z_2)\,,
\end{equation}
with $y=\lbrace v,w,z_1,z_2\rbrace$ being the variables that parameterize the $SU(3)_C\times SU(2)_W\times U(1)_Y$ gauge group. The characters $\chi_a$ for the single particle modules of the spurions are a composition of the characters for the conformal and gauge group representation $\mathbf{R}$ of the spurions,
\begin{equation}
\chi_a (\mathcal{D} ; x, y) = \chi_{[\Delta_a , l_a]} (\mathcal{D}; x ) \cdot \chi_{\mathbf{R}_a}^{U(1)_Y}(v)\cdot \chi_{\mathbf{R}_a}^{SU(2)_W}(w)\cdot \chi_{\mathbf{R}_a}^{SU(3)_C}(z_1,z_2)\,.
\end{equation}
The explicit form of the group measures and characters is given in appendix~\ref{app:ConfChar}.
Note that in order to fully describe the flavor structure of the SM we would have to work with three independent instances of each fermion to implement the three fermion generations. This would increase the number of terms in the generating function at each order exponentially. However, we can still get some information about the number of invariants with $N_f$ flavors by simply suppressing the flavor indices and adding the same fermion spurion $N_f$ times, i.e.~we can write the complete PE as
\begin{equation}
\text{PE} \bigg[\bigg\lbrace \frac{\phi_a}{\mathcal{D}^{\Delta_a}} \bigg\rbrace \bigg]= \prod_b \text{PE} \bigg[\bigg\lbrace \frac{\phi_b}{\mathcal{D}^{\Delta_b}} \bigg\rbrace \bigg] \prod_f \text{PEF} \bigg[\bigg\lbrace \frac{\phi_f}{\mathcal{D}^{\Delta_f}} \bigg\rbrace \bigg]^{N_f}\,,
\end{equation}
where the index $b$ runs over all bosons and $f$ over all fermions.
Next we expand the Hilbert series according to the mass dimension of the operators. We will neglect all pure SM contributions to the Hilbert series, which are given in \cite{Henning:2015alf}. The first gravity operators appear at dimension 6, the Hilbert series being
\begin{equation}
\mathcal{H}_6 = C_L^3+C_R^3 + B_L^2 C_L+B_R^2 C_R+H C_L^2 H^{\dagger }+H C_R^2 H^{\dagger }+C_L G_L^2+C_R G_R^2+C_L W_L^2+C_R W_R^2 \,.
\end{equation}
This includes the pure gravity contributions discussed in section~\ref{subsec:gravityVacuum} plus mixed SM-gravity terms. Note that at this mass dimension, the latter operators only contain SM bosons. An explicit operator basis is given by

\begin{eqnarray}
\mathcal{L}_6 &=&  \frac{c_1}{\Lambda^2} C_{\mu\nu}\,^{\rho\sigma} C^{\mu\nu\alpha\beta} C_{\alpha\beta\rho\sigma} + \frac{\tilde{c}_1}{\Lambda^2}C_{\mu\nu}\,^{\rho\sigma} C^{\mu\nu\alpha\beta} \tilde{C}_{\alpha\beta\rho\sigma} \nonumber \\
&&\! +\, \frac{c_2}{\Lambda^2} H^\dagger H  C_{\mu\nu\rho\sigma} C^{\mu\nu\rho\sigma}  + \frac{\tilde{c}_2}{\Lambda^2} H^\dagger H  C_{\mu\nu\rho\sigma} \tilde{C}^{\mu\nu\rho\sigma} \nonumber \\
&&\! +\, \frac{c_3}{\Lambda^2} B^{\mu\nu} B^{\rho\sigma} C_{\mu\nu\rho\sigma} + \frac{\tilde{c}_3}{\Lambda^2} B^{\mu\nu} B^{\rho\sigma} \tilde{C}_{\mu\nu\rho\sigma} + \frac{c_4}{\Lambda^2} G^{\mu\nu} G^{\rho\sigma} C_{\mu\nu\rho\sigma}+ \frac{\tilde{c}_4}{\Lambda^2} G^{\mu\nu} G^{\rho\sigma} \tilde{C}_{\mu\nu\rho\sigma} \nonumber \\
&&\! +\, \frac{c_5}{\Lambda^2} W^{\mu\nu} W^{\rho\sigma} C_{\mu\nu\rho\sigma} + \frac{\tilde{c}_5}{\Lambda^2} W^{\mu\nu} W^{\rho\sigma} \tilde{C}_{\mu\nu\rho\sigma} \,.
\label{eq:dim6GRSMEFT}
\end{eqnarray}
There are no new gravity operators at mass dimension 7. However, there is a multitude of terms in the Hilbert series at mass dimension 8. This is the first order where operators with SM fermions appear. For one flavor, i.e.~$N_f=1$, the part of the Hilbert series that involves gravity reads
\begin{eqnarray}
\mathcal{H}_8 &=& C_L^4+H H^{\dagger } C_L^3+H^2 \left(H^{\dagger }\right)^2 C_L^2+2 B_L^2 C_L^2+B_R^2 C_L^2+C_R^2 C_L^2+2 G_L^2 C_L^2+G_R^2 C_L^2+2 W_L^2 C_L^2\nonumber\\
&&\! +\,W_R^2 C_L^2 +H Q u C_L^2+H \mathcal{D}^2 H^{\dagger } C_L^2+e L H^{\dagger } C_L^2+d Q H^{\dagger } C_L^2+H e^{\dagger } L^{\dagger } C_L^2+H d^{\dagger } Q^{\dagger } C_L^2\nonumber\\
&&\! +\,H^{\dagger } Q^{\dagger } u^{\dagger } C_L^2+d e u^2 C_L +H H^{\dagger } B_L^2 C_L +H H^{\dagger} G_L^2 C_L+B_L G_L^2 C_L +H H^{\dagger } W_L^2 C_L\nonumber\\
&&\! +\,B_L W_L^2 C_L+d Q^2 u C_L+H Q \mathcal{D}^2 u C_L+e L Q u C_L +e L \mathcal{D}^2 H^{\dagger } C_L +d Q \mathcal{D}^2 H^{\dagger } C_L\nonumber\\
&&\! +\,H Q u B_L C_L+d \mathcal{D} d^{\dagger } B_L C_L+e \mathcal{D} e^{\dagger } B_L C_L+H \mathcal{D}^2 H^{\dagger } B_L C_L+e L H^{\dagger } B_L C_L+d Q H^{\dagger } B_L C_L\nonumber\\
&&\! +\,L \mathcal{D} L^{\dagger } B_L C_L+Q \mathcal{D} Q^{\dagger } B_L C_L+\mathcal{D} u u^{\dagger } B_L C_L+H Q u G_L C_L+d \mathcal{D} d^{\dagger } G_L C_L+d Q H^{\dagger } G_L C_L\nonumber\\
&&\! +\,Q \mathcal{D} Q^{\dagger } G_L C_L +\mathcal{D} u u^{\dagger } G_L C_L+H Q u W_L C_L+H \mathcal{D}^2 H^{\dagger } W_L C_L+e L H^{\dagger} W_L C_L\nonumber\\
&&\! +\,d Q H^{\dagger } W_L C_L+L \mathcal{D} L^{\dagger } W_L C_L +Q \mathcal{D} Q^{\dagger } W_L C_L+H H^{\dagger } B_L W_L C_L+C_R^4+H H^{\dagger } C_R^3\nonumber\\
&&\! +\,H^2 \left(H^{\dagger }\right)^2 C_R^2+B_L^2 C_R^2+2 B_R^2 C_R^2+H Q u C_R^2 +H \mathcal{D}^2 H^{\dagger } C_R^2+e L H^{\dagger } C_R^2+d Q H^{\dagger } C_R^2\nonumber\\
&&\! +\,H e^{\dagger } L^{\dagger } C_R^2+H d^{\dagger } Q^{\dagger } C_R^2+H^{\dagger } Q^{\dagger } u^{\dagger } C_R^2+C_R^2 G_L^2+2 C_R^2 G_R^2 +H H^{\dagger } C_R G_R^2+B_R C_R G_R^2\nonumber\\
&&\! +\,C_R^2 W_L^2+2 C_R^2 W_R^2+H H^{\dagger } C_R W_R^2+B_R C_R W_R^2 + d^{\dagger } e^{\dagger } \left(u^{\dagger }\right)^2 C_R+H H^{\dagger } B_R^2 C_R\nonumber\\
&&\! +\,H \mathcal{D}^2 e^{\dagger } L^{\dagger } C_R+H \mathcal{D}^2 d^{\dagger } Q^{\dagger } C_R+d^{\dagger } \left(Q^{\dagger }\right)^2 u^{\dagger } C_R+\mathcal{D}^2 H^{\dagger } Q^{\dagger } u^{\dagger } C_R+e^{\dagger } L^{\dagger } Q^{\dagger } u^{\dagger } C_R\nonumber\\
&&\! +\,d \mathcal{D} d^{\dagger } B_R C_R+e \mathcal{D} e^{\dagger } B_R C_R +H \mathcal{D}^2 H^{\dagger } B_R C_R+L \mathcal{D} L^{\dagger } B_R C_R+H e^{\dagger } L^{\dagger } B_R C_R\nonumber\\
&&\! +\,Q \mathcal{D} Q^{\dagger } B_R C_R+H d^{\dagger } Q^{\dagger } B_R C_R+\mathcal{D} u  u^{\dagger } B_R C_R+H^{\dagger } Q^{\dagger } u^{\dagger } B_R C_R+d \mathcal{D} d^{\dagger } C_R G_R\nonumber\\
&&\! +\,Q \mathcal{D} Q^{\dagger } C_R G_R+H d^{\dagger } Q^{\dagger } C_R G_R+\mathcal{D} u u^{\dagger } C_R G_R+H^{\dagger } Q^{\dagger } u^{\dagger } C_R G_R+H \mathcal{D}^2 H^{\dagger } C_R W_R\nonumber\\
&&\! +\,L \mathcal{D} L^{\dagger} C_R W_R+H e^{\dagger } L^{\dagger } C_R W_R+Q \mathcal{D} Q^{\dagger } C_R W_R+H d^{\dagger } Q^{\dagger } C_R W_R+H^{\dagger } Q^{\dagger }
 u^{\dagger } C_R W_R\nonumber\\
&&\! +\,H H^{\dagger } B_R C_R W_R \,.
\end{eqnarray}
A classification of the dimension 8 operators of our basis is given in Table~\ref{tab:dim8classification}, while an explicit form for all 103 of them for $N_f=1$ can be found in Tables~\ref{tab:dim8Boson} and \ref{tab:dim8Ferm} of appendix~\ref{app:GRSMEFT8}.%
\footnote{As a curiosity, we find a single dimension-8 operator, $\mathcal{O}_{uedC}$ in Table~\ref{tab:dim8Ferm}, that violates baryon and lepton numbers, by $\Delta B = \Delta L = 1$. }
\begin{table}[t!]
\centering
\begin{tabular}{ l | *4c }    
Structure 			& $N_f$ & $N_f = 1$ & $N_f=3$ & Representative Operator \\\hline
$C^4$			&	3			&	3	&	3	&	$(C_{\mu\nu\rho\sigma} C^{\mu\nu\rho\sigma})^2$ \\
$C^3 H^2$		&	2			&	2	&	2	&	$H^\dagger H (C_{\mu\nu}\,^{\rho\sigma} C^{\mu\nu\alpha\beta} C_{\alpha\beta\rho\sigma})$\\
$C^2 H^4$		&	2			&	2	&	2	&	$(H^\dagger H)^2 (C_{\mu\nu\rho\sigma} C^{\mu\nu\rho\sigma})$\\
$C^2 X^2$		&	18			&	18	&	18	&	$B_{\mu\nu} B^{\rho\sigma} C^{\mu\nu\alpha\beta} C_{\alpha\beta\rho\sigma}$ \\
$C H^2 X^2$		&	8			&	8	&	8	&	$H^\dagger H (C^{\mu\nu\rho\sigma} W^a_{\mu\nu} W^a_{\rho\sigma})$\\
$C X^3$			&	4			&	4	&	4	&	$C^{\mu\nu\rho\sigma} W^a_{\mu\nu} W^a_{\rho\alpha} B^\alpha\,_{\sigma}$\\
$C^2 H \psi^2$		&	$12 N_f^2$	&	12	&	108	&	$\bar{Q}_L H d_R (C_{\mu\nu\rho\sigma} C^{\mu\nu\rho\sigma})$\\
$C H X \psi^2$		&	$16 N_f^2$	&	16	&	144	&	$C_{\mu\nu\rho\sigma} (\bar{Q}_L \sigma^{\mu\nu} d_R) \tau^a H\, W^{a,\, \rho\sigma}$\\
$C \psi^4$		&	$\tfrac{N_f^2}{3} (17 N_f^2 +3 N_f -2)$	&	6	&	480	&	$\epsilon_{jk} C_{\mu\nu\rho\sigma} (\bar{Q}_L^j \sigma^{\mu\nu} u_R) (\bar{L}^k_L \sigma^{\rho\sigma} e_R)$\\\hline
$C X \psi^2 \mathcal{D}$	&	$20 N_f^2$	&	20	&	180	&	$C_{\mu\nu\rho\sigma} (\bar{Q}_L \gamma^{\mu}\tau^a \nabla^\nu Q_L) W^{a,\, \rho\sigma}$\\\hline
$C^2 H^2 \mathcal{D}^2$	&	2	&	2	&	2	&	$(\nabla_\mu H)^\dagger (\nabla^\mu H) (C_{\mu\nu\rho\sigma} C^{\mu\nu\rho\sigma})$\\
$C H^2 X \mathcal{D}^2$	&	4	&	4	&	4	&	$C_{\mu\nu\rho\sigma} (\nabla^\mu H)^\dagger \tau^a (\nabla^\nu H) W^{a,\, \rho\sigma}$\\
$C H \psi^2 \mathcal{D}^2$	&	$6 N_f^2$	&	6	&	54	&	$C_{\mu\nu\rho\sigma} (\bar{Q}_L \sigma^{\mu\nu} \nabla^\rho d_R) \nabla^\sigma H$\\\hline
Total				&	$43+ \tfrac{N_f^2}{3} (17 N_f^2 + 3N_f+160)$	&	103	&	1009	&	\\
\end{tabular}
\caption{Classification of dimension-8 operators containing gravity interactions. $C$ denotes the Weyl tensor, whereas $H$, $\psi$, $X$ and $\mathcal{D}$ stand for the Higgs, fermions, gauge fields and derivatives respectively. We show the number of operators in each class for $N_f$ fermion flavors and give one exemplary operator for each class.}
\label{tab:dim8classification}
\end{table}
\subsection{Comments on the GRSMEFT operator basis}
Let us comment on some interesting aspects of the GRSMEFT operator basis. 

Matching what can be derived based on little group covariance and locality of on-shell massless 3-particle amplitudes (see e.g.~\cite{Elvang:2013cua}), we find in our basis the corresponding EFT operators modifying gravitational trilinear vertices: $C_{\mu\nu}\,^{\rho\sigma} C^{\mu\nu\alpha\beta} \overset{\text{\tiny($\sim$)}}{C}_{\alpha\beta\rho\sigma}$ (3 gravitons), $X^{\mu\nu} X^{\rho\sigma} \overset{\text{\tiny($\sim$)}}{C}_{\mu\nu\rho\sigma}$ (1 graviton and 2 gauge bosons), and $H^\dagger H  C^{\mu\nu\rho\sigma} \overset{\text{\tiny($\sim$)}}{C}_{\mu\nu\rho\sigma}$ (2 gravitons and 1 scalar, once the Higgs gets a vacuum expectation value). This one-to-one correspondence follows from the fact that our basis does not include terms that vanish on the free EOMs.

Similar to the EFT of pure gravity in vacuum, the leading dimension-4 Lagrangian induces, at one loop, RG evolution for some of the operators in the GRSMEFT, although in our basis all such operators involve SM fields only~\cite{tHooft:1973bhk,tHooft:1974toh,Deser:1974cz,Deser:1974xq,Deser:1974cy}. In this regard, the absence of (one-loop) divergences associated with mixed SM-gravity operators, such as $X^{\mu\nu} X^{\rho\sigma} C_{\mu\nu\rho\sigma}$, can be understood, in the particular case of gravity coupled to a $U(1)$ gauge field,
by the invariance of the leading Einstein-Maxwell Lagrangian under vector field duality transformation~\cite{Deser:1975sx}, or from supersymmetry in the case of the Einstein-Yang-Mills system~\cite{Deser:1981fk}. To rederive these non-renormalization results from helicity selection rules as in~\cite{Cheung:2015aba} would certainly be interesting~\cite{workinprogress}.
Heavy (charged) matter fields give finite contributions~\cite{Berends:1975ah,Drummond:1979pp,Bastianelli:2008cu} to the operators of the GRSMEFT, e.g.~a Dirac fermion of unit hypercharge and mass $\Lambda$ generates $c_3 = -\tfrac{1}{90} (g'/4\pi)^2$, as well as contributions to several other operators in the SMEFT. In this regard, one should note that the latter operators, for instance $(B_{\mu \nu}B^{\mu \nu})^2$, receive direct contributions from the heavy dynamics, of $\mathcal{O}(g'^4/\Lambda^4)$, as well as contributions from operators with $R_{\mu \nu}$, $R$, which when rewritten in our basis are relatively suppressed by powers of $\Lambda/g'\Mpl$. We note in passing that we have not found in the literature the corresponding calculation for the coefficients of the dimension-6 Higgs-gravity operators in Eq.~(\ref{eq:dim6GRSMEFT}).

One can power-count, as in Eq.~(\ref{GRNDA}) for the pure gravity EFT, the size of the Wilson coefficients in a generic GRSMEFT. Focussing for simplicity on the subclass of operators involving gauge fields and gravity
\begin{equation}
\mathcal{L}_{eff} = \frac{m_*^4}{g_*^2} L \! \left( \frac{R_{\mu\nu\rho\sigma}}{m_*^2}, \frac{\nabla_\mu}{m_*}, \frac{\epsilon X_{\mu \nu}}{m_*^2}\right) - \frac{\hat M_{\rm{pl}}^2}{2} R - \frac{1}{4 \hat g^2} X_{\mu \nu} X^{\mu \nu}\,,
\label{GRSMNDA}
\end{equation}
where we introduced a ``fundamental'' kinetic term for the gauge field, with coupling $\hat g \lesssim g_*$, $\epsilon$ parametrizes the (multipole) charge of the heavy states that have been integrated out, and $\nabla_\mu = \partial_\mu + i \omega_\mu + i \epsilon' X_\mu$ with $\epsilon'$ parametrizing the (monopole) charge of the particles, if any, that remain in the EFT, fixing then the low-energy gauge coupling to $g = \epsilon' \hat g$~\cite{Liu:2016idz}. 
From Eq.~(\ref{GRSMNDA}) one can conclude that there could be situations in which gravitational operators such as $X^{\mu\nu} X^{\rho\sigma} C_{\mu\nu\rho\sigma}$ are enhanced compared to non-gravitational ones like $(X_{\mu\nu} X^{\mu\nu})^2$, e.g.~if $\epsilon \ll \epsilon'$, a pattern that arises for instance from milli-charged particles -- an axion would belong to this category. This however does not appear as an optimal (phenomenological) scenario, since the light charged SM particles that remain in the spectrum below $m_*$, e.g.~the electron, would dominate the new EFT coefficients (since $\epsilon' \gg \epsilon$) after being themselves integrated out at even lower energies. This is unless there exists no charged particle below $m_*$, i.e.~$m_* \ll m_e$, for which, while $\epsilon \ll 1$, $\epsilon' = 0$ -- for the GRSMEFT, this would mean a very low new physics scale, yet with an interesting and unexplored parameter space in terms of mixed SM-gravity effects. In scenarios where $\epsilon \gg \epsilon' \neq 0$, the non-gravitational operators are instead comparatively enhanced. 
We also note that from a purely low-energy point of view, there seems to be nothing wrong with taking mixed SM-gravity operators, such as $X^{\mu\nu} X^{\rho\sigma} C_{\mu\nu\rho\sigma}$, of size $\mathcal{O}(g^2/g_*^2m_*^2)$, as the leading deformation in the EFT, in the sense that quantum corrections within the EFT do not point towards large non-gravitational operators as long as the cutoff, which saturates the loops in the UV, satisfies $m_* \lesssim 4 \pi (g_* / g) \Mpl$, and this even for $g_* \ll g$, although such a condition goes against NDA.%
\footnote{However, at least for the mixed SM-gravity operator $X^{\mu\nu} X^{\rho\sigma} C_{\mu\nu\rho\sigma}$ with $X_{\mu \nu}$ associated with a $U(1)$ gauge field, this possibility seems to be in tension with arguments related to the weak gravity conjecture~\cite{ArkaniHamed:2006dz,Bellazzini:2019xts}.}

Regardless of these facts, there always remains the obstacle that to probe operators intrinsically sensitive to gravitational physics, by which we mean those that do not depend on $R_{\mu \nu}$, $R$ and therefore do not contribute to the SMEFT, one needs to overcome the $\Mpl$ suppression that comes with gravitational interactions. This is of course the reason why experimental constraints on the SMEFT are much more stringent than those on the rest of the GRSMEFT. The question of how to test gravitational EFT operators has been partly investigated before, e.g.~in~\cite{Drummond:1979pp} for the Einstein-Maxwell system after integrating out the electron, or more recently in~\cite{Goon:2016une} for more general situations yet concentrating still on photon propagation around non-trivial gravitational backgrounds. For purely gravitational operators, \cite{Endlich:2017tqa} studied their effects on the gravitational waves from merging black holes. In all these situations, the conclusion is that for the effects of the higher-dimensional operators to be observable, the typical size (e.g.~the Schwarzschild radius) and distance from the gravitational source should be of the order of the (inverse) cutoff of the EFT. Leaving aside our preconceptions on the expected size of the mixed SM-gravity operators with respect to non-gravitational ones, one should consider probing the former at high-energy colliders~\cite{workinprogress}.
Finally, we think it is worthwhile to further investigate and extend the theoretical constraints based on causality, unitarity and analyticity to the full set of operators in Eq.~(\ref{eq:dim6GRSMEFT}).


\section{Summary}\label{sec:summary}

In this paper we have developed a systematic methodology to construct operators bases for relativistic EFTs with gravity. Our approach relies on Hilbert series and conformal representation theory, and makes use of the Weyl tensor as basic building block of gravitational operators.

We applied our method to build several compelling EFTs: pure gravity, a shift-symmetric scalar coupled to gravity, and the GRSMEFT, i.e.~gravity coupled to the SM of particle physics. Needless to say, the same techniques could be used as well to construct other gravitational EFTs of interest. Along the way, we reviewed several important aspects of the EFT of gravity in vacuum, identified several operator redundancies of the shift-symmetric scalar EFT, and explored a few salient features of the set of potential deformations of the SM coupled to gravity.

Finally, we recall that while GR is one of the most solid theories in fundamental physics, its deformations remain largely unconstrained, in particular at distances where the other fundamental forces operate on. It is our hope that the results we have obtained in this paper, specially for what regards the GRSMEFT, will contribute to improving our knowledge of these issues, which we believe are of great theoretical and phenomenological interest.


\vspace{1cm}
{\bf Acknowledgments} 
\vspace{0.5cm}

We would like to thank Brian Henning for insightful comments on the manuscript, Brando Bellazzini, Maximilian Dichtl and Enrico Trincherini for useful discussions, as well as Hitoshi Murayama for a stimulating seminar at TUM that led us to think about Hilbert series for EFTs.  The work of MR, JS and AW has been partially supported by the DFG Cluster of Excellence 2094 ``Origins'', by the Collaborative Research Center SFB1258 and BMBF grant no.~05H18WOCA1. MR is supported by the Studienstiftung des deutschen Volkes.


\newpage

\appendix

\section{Group Characters}\label{app:ConfChar}

In this appendix we summarize the Haar integration measures and group characters, taken from~\cite{Hanany:2008sb}, that were used to derive our main results. 


\subsection{Integration Measures}

The Haar integration measures over the SM gauge groups can be written as contour integrals in the complex plane of the variables parametrizing the groups
\begin{align}
\int d\mu_{U(1)_Y}(v) &= \frac{1}{2\pi i}\oint_{|v | = 1} \frac{dv}{v}\,,\\
\int d\mu_{SU(2)_W}(w) &= \frac{1}{2\pi i} \oint_{|w| = 1} \frac{dw}{w} \big(1-w^2\big) \,,\\
\int d\mu_{SU(3)_C}(z_1, z_2) &= \frac{1}{(2\pi i)^2} \oint_{|z_1| = 1} \oint_{|z_2| = 1} \frac{dz_1}{z_1}\frac{dz_2}{z_2} \bigg( 1- z_1 z_2\bigg) \bigg( 1 - \frac{z_1^2}{z_2}\bigg) \bigg( 1 - \frac{z_2^2}{z_1}\bigg) \,.
\end{align}
Note that these expressions differ from the ones in~\cite{Henning:2015alf}, since the Haar measures that we use involve only the positive roots and therefore have no Weyl group normalization. This simplified measure can be used when integrating over class functions, i.e.~functions $f(g)$ which satisfy $f(hgh^{-1})=f(g)$ for $h,g\in G$, since they are invariant under the Weyl group. Note that all characters are class functions.
For the integration measure over the euclidean Lorentz group $SO(4)\simeq SU(2)_L\otimes SU(2)_R$ we use
\begin{equation}
\int d\mu_{\rm Lorentz}(x) = \int d\mu_{SU(2)_L\otimes SU(2)_R}(x) = \frac{1}{(2\pi i)^2} \oint_{|x_1| = 1} \oint_{|x_2| = 1} \frac{dx_1}{x_1}\frac{dx_2}{x_2} \big(1-x_1^2\big) \big(1-x_2^2\big)\,,
\end{equation}
where $x=\lbrace x_1 ,x_2\rbrace$.


\subsection{Characters for SM Gauge Representations}

The characters for all gauge group representations appearing in the SM are given by
\begin{gather}
\chi_{\mathbf Q}^{U(1)_Y} (v) = v^Q\,,\\
\chi_{\mathbf 2}^{SU(2)_W} (w) = \chi_{\mathbf{\bar{2}}}^{SU(2)_W} (w)= w + \frac{1}{w}\,,\quad \chi_{\mathbf{adj}}^{SU(2)_W} (w) = w^2 + 1 + \frac{1}{w^2}\,,\\
\chi_{\mathbf 3}^{SU(3)_C} (z_1,z_2) = z_1 + \frac{z_2}{z_1} + \frac{1}{z_2}\,,\quad \chi_{\mathbf{\bar{3}}}^{SU(3)_C} (z_1,z_2) = z_2 + \frac{z_1}{z_2} + \frac{1}{z_1}\,, \nonumber \\
\chi_{\mathbf{adj}}^{SU(3)_C} (z_1,z_2) = z_1 z_2 +\frac{z_2^2}{z_1} + \frac{z_1^2}{z_2} + 2 + \frac{z_1}{z_2^2} + \frac{z_2}{z_1^2} + \frac{1}{z_1z_2}\,.
\end{gather}
Characters for the Lorentz group are products of $SU(2)$ characters
\begin{equation}
\chi_{(l_1,l_2)} (x) = \chi^{SU(2)_L}_{l_1}(x_1) \cdot \chi^{SU(2)_R}_{l_2}(x_2)\,,
\end{equation}
with
\begin{align}
\chi_{1/2}(x) &= x + \frac{1}{x}\,,\qquad & &\chi_1 (x) = x^2 + 1 +\frac{1}{x^2}\,,\nonumber \\
\chi_{3/2}(x) &= x^3 + x + \frac{1}{x} + \frac{1}{x^3}\,,& &\chi_2 (x) = x^4 + x^2 +1 + \frac{1}{x^2} + \frac{1}{x^4}\,.
\end{align}


\subsection{Conformal Characters}

The characters for all unitary conformal representations we use in this work are given by~\cite{Henning:2015alf,Henning:2017fpj,Barabanschikov:2005ri}
\begin{align}
\chi_{[0,(0,0)]} (\mathcal{D}; x) &= \mathcal{D}\, P(\mathcal{D} ; x ) (1 - \mathcal{D}^2)\,,\\
\chi_{[3/2,(1/2,0)]} (\mathcal{D}; x ) &= \mathcal{D}^{\frac{3}{2}}\, P(\mathcal{D} ;x ) \big( \chi_{(1/2,0)}(x) - \mathcal{D}\, \chi_{(0,1/2)}(x)\big)\,,\\
\chi_{[3/2,(0,1/2)]} (\mathcal{D}; x ) &= \mathcal{D}^{\frac{3}{2}}\, P(\mathcal{D} ; x ) \big( \chi_{(0,1/2)}(x) - \mathcal{D}\, \chi_{(1/2,0)}(x)\big)\,,\\
\chi_{[2,(1,0)]} (\mathcal{D};x ) &= \mathcal{D}^{2}\, P(\mathcal{D} ; x ) \big( \chi_{(1,0)}(x) - \mathcal{D}\, \chi_{(1/2,1/2)}(x) + \mathcal{D}^2\big)\,,\\
\chi_{[2,(0,1)]} (\mathcal{D}; x ) &= \mathcal{D}^{2}\, P(\mathcal{D} ; x ) \big( \chi_{(0,1)}(x) - \mathcal{D}\, \chi_{(1/2,1/2)}(x) + \mathcal{D}^2\big)\,,\\
\chi_{[3,(2,0)]} (\mathcal{D}; x) &= \mathcal{D}^{3}\, P(\mathcal{D} ; x) \big( \chi_{(2,0)}(x) - \mathcal{D}\, \chi_{(3/2,1/2)}(x) + \mathcal{D}^2 \chi_{(1,0)}(x) \big)\,,\\
\chi_{[3,(0,2)]} (\mathcal{D}; x) &= \mathcal{D}^{3}\, P(\mathcal{D} ; x) \big( \chi_{(0,2)}(x) - \mathcal{D}\, \chi_{(1/2,3/2)}(x) + \mathcal{D}^2 \chi_{(0,1)}(x) \big)\,,
\end{align}
with the momentum generating function $P(\mathcal{D} ; x)$ \cite{Henning:2015alf}
\begin{equation}
P(\mathcal{D} ; x) = \frac{1}{(1-\mathcal{D} x_1 x_2)(1-\mathcal{D}/(x_1 x_2))(1-\mathcal{D} x_1/x_2)(1-\mathcal{D} x_2 / x_1)}\,.
\end{equation}


\section{Operator Redundancies}\label{app:redundOp}

In Section~\ref{subsec:ScalarShiftSymmetric} we identified two redundant operators in the basis of~\cite{Solomon:2017nlh} for a shift-symmetric scalar coupled to gravity. Here we show how these can be related to the operator basis in Eq.~(\ref{eq:6derScalarBasis}). Dropping freely all terms proportional to the free EOM, i.e.~any terms containing $R_{\mu\nu}$, $R$ or $\nabla^\mu C_{\mu\nu\rho\sigma}$, the first operator can be rewritten as

\begin{eqnarray}
(\nabla_\alpha R_{\mu\nu\rho\sigma})(\nabla^\alpha R^{\mu\nu\rho\sigma}) &=& - C_{\mu\nu\rho\sigma} \nabla^2 C^{\mu\nu\rho\sigma} = C_{\mu\nu\rho\sigma}( 4\, C^{\lambda\nu\rho\alpha} C^\mu\,_\lambda\,^\sigma\,_\alpha  + C^{\lambda\alpha\rho\sigma} C^{\mu\nu}\,_{\lambda\alpha} )\nonumber \\
&=& 3\, C_{\mu\nu}\,^{\rho\sigma} C^{\mu\nu\alpha\beta} C_{\alpha\beta\rho\sigma} = 3\, \mathcal{O}_3\,,
\end{eqnarray}
where we used IBP in the first step and Eq.~(\ref{eq:boxWeyl}) in the second. Since there is only one independent CP even Riemann invariant with three Riemann tensors in four dimensions~\cite{Fulling:1992vm}, it is clear that the first line is proportional to $\mathcal{O}_3$. To find the exact relation one has to use dimensionally dependent identities, which can be conveniently implemented with the Invar package~\cite{MartinGarcia:2008qz}.
Again throwing away all terms that vanish due to the free EOM, the second operator can be rewritten as
\begin{eqnarray}
R_{\mu\nu\alpha\beta}(\nabla^\mu\nabla^\alpha\phi)(\nabla^\nu\nabla^\beta\phi) &=& - C_{\mu\nu\alpha\beta}(\nabla^\alpha\phi)(\nabla^\mu\nabla^\nu\nabla^\beta\phi) \nonumber \\
&=& -\frac{1}{2}C_{\mu\nu\alpha\beta}(\nabla^\alpha\phi)((\lbrace \nabla^\mu,\nabla^\nu\rbrace + [\nabla^\mu,\nabla^\nu])\nabla^\beta\phi)\nonumber \\
&=& -\frac{1}{2}C_{\mu\nu\beta\alpha} C^{\mu\nu\beta}\,_\sigma (\nabla^\alpha\phi)(\nabla^\sigma\phi ) \nonumber \\
&=& -\frac{1}{8}(C_{\alpha\beta\rho\sigma})^2 (\nabla_\mu\phi )^2\ = -\frac{1}{8} \mathcal{O}_4\,.
\end{eqnarray}
In the first step we used IBP and wrote the covariant derivatives in the last parenthesis as the sum of its commutator and anti-commutator. The term with the anti-commutator vanishes, since it is contracted with $C_{\mu\nu\alpha\beta}$, which is antisymmetric under $\mu\leftrightarrow \nu$. The commutator yields a Riemann tensor, which after removing the $R_{\mu\nu}$ and $R$ components coincides with the Weyl tensor. In the last step we made use of the identity~\cite{Edgar:2001vv}
\begin{equation}
C_{\mu\nu\beta\alpha} C^{\mu\nu\beta}\,_\sigma = \frac{1}{4} g_{\alpha\sigma} C_{\mu\nu\gamma\delta} C^{\mu\nu\gamma\delta}\,.
\end{equation}
%


\section{Plethystic Exponential}\label{app:PE}

In Section~\ref{subsec:HS} we introduced the (fermionic) PE as the generating function for the characters of (anti-)symmetric tensor products~\cite{Benvenuti:2006qr,Feng:2007ur,Hanany:2014dia}. In the following we give a sketchy derivation to justify the form of the PE which we use here. Readers looking for mathematical rigour should refer to commutative algebra textbooks, such as~\cite{Neusel,Sturmfels}.


\subsection{Bosonic Plethystic Exponential}

We want to compute the sum over the characters of all symmetric tensor products of a representation $\mathbf{R}$ weighted by a spurion $q$, i.e. 
\begin{equation}
\sum_{d=0}^\infty q^d\,  \chi_{\text{Sym}^d(\mathbf{R})} (g)\,.
\end{equation}
For $g\in G$, let $\mathbf{R}_V(g) \in GL(V)$ be the linear action of the group element $g$ on a $n$-dimensional vector space $V$, i.e.~$\mathbf{R}$ is a $n$ dimensional group representation. Now let us assume that $\mathbf{R}_V(g)$ can be diagonalized. We take its set of eigenvectors, i.e.~$\lbrace e_1,\ldots ,e_n\rbrace$ with $\mathbf{R}_V (g) e_i = \lambda_i e_i$, as a basis for $V$. In this basis the group character is given by the sum over the eigenvalues $\chi_{\mathbf{R}}(g) = $Tr$(\mathbf{R}_V(g))= \sum_{i=1}^n \lambda_i$.
The symmetric tensor product $\text{Sym}^d(\mathbf{R})$ is the action of the group element $g$ on the symmetric tensor product of the vector space $V$, i.e.~$\text{Sym}^d(V)$. We denote this linear map by $\mathbf{R}^{\otimes d}_{\text{Sym}^d(V)}(g)\in GL($Sym$^d(V))$.%
\footnote{This map is the tensor product representation $\otimes^d \mathbf{R}$ acting on $\text{Sym}^d(\mathbf{V})$.} 
A simple basis for $\text{Sym}^d(V)$ is $\lbrace \tfrac{1}{d!} \sum_{\sigma\in S_d} e_{i_{\sigma(1)}}\otimes \ldots \otimes e_{i_{\sigma(d)}} | 1\leq i_1\leq \ldots \leq i_d\leq n\rbrace$, where $S_d$ is the symmetric group. As an explicit example let us write down the basis for Sym$^2(V)$ and  dim$(V)=3$
\begin{equation}
\lbrace e_1 \otimes e_1\,, e_2\otimes e_2\,, e_3\otimes e_3\,, \tfrac{1}{2} (e_1\otimes e_2 + e_2 \otimes e_1)\,, \tfrac{1}{2} (e_1\otimes e_3 + e_3 \otimes e_1)\,, \tfrac{1}{2} (e_2\otimes e_3 + e_3 \otimes e_2) \rbrace\,.\\
\end{equation}
$\chi_{\text{Sym}^2(\mathbf{R})}(g)$ is obtained by summing over the eigenvalues corresponding to these basis elements
\begin{equation}
\chi_{\text{Sym}^2(\mathbf{R})}(g) = \text{Tr}\big(\mathbf{R}^{\otimes 2}_{\text{Sym}^2(V)}(g)\big) = \lambda_1^2 + \lambda_2^2 +\lambda_3^2 + \lambda_1\lambda_2 + \lambda_1\lambda_3 + \lambda_2\lambda_3 = \tfrac{1}{2} (\chi_{\mathbf{R}}(g)^2 + \chi_{\mathbf{R}}(g^2))\,,
\label{eq:charSymSq}
\end{equation}
where the trace is a regular matrix trace. In Eq.~(\ref{eq:charSymSq}) we also verified the symmetric square formula for characters that we already found in the explicit example for the Hilbert Series in Eq.~(\ref{eq:PEexample}). For general $n$ and $d$ the character can be written as
\begin{equation}
\chi_{\text{Sym}^d(\mathbf{R})}(g) = \text{Tr}\big(\mathbf{R}^{\otimes d}_{\text{Sym}^d(V)}(g)\big) = \sum_{i_1 +i_2+\ldots +i_n =d} \lambda_1^{i_1}\lambda_2^{i_2}\cdots \lambda_n^{i_n}\,,
\label{eq:charSymd}
\end{equation}
where the sum is over all partitions $\lbrace i_1,i_2,\ldots ,i_n\rbrace $ with $i_1 +i_2+\ldots +i_n =d$. The $i_k$ indicate the number of times $e_k$ appears in the corresponding basis element and therefore also the power of $\lambda_k$ in its eigenvalue. Each partition corresponds to a basis element of $\text{Sym}^d(V)$. This is easily seen in our example with $d=2$ and $n=3$ with the partitions being $2 = 2+ 0 + 0 = 0 + 2 + 0 = 0 + 0 +2 = 1 + 1 + 0 = 1 + 0 + 1 = 0 + 1 + 1$. Summing over $d$ in Eq.~(\ref{eq:charSymd}) yields the generating function
\begin{eqnarray}
\sum_{d=0}^\infty \chi_{\text{Sym}^d(\mathbf{R})}(g)\, q^d &=& \sum_{d=0}^\infty  q^d \sum_{i_1 +i_2+\ldots +i_n =d} \lambda_1^{i_1}\lambda_2^{i_2}\cdots \lambda_n^{i_n} = \bigg( \sum_{i_1=0}^\infty (\lambda_1 q)^{i_1}\bigg)\cdots \bigg( \sum_{i_n=0}^\infty (\lambda_n q)^{i_n}\bigg) \nonumber \\
&=& \frac{1}{\prod_{i=1}^n (1-\lambda_i q)} = \frac{1}{\text{det}(\mathbbm{1}-\mathbf{R}_V(g)\, q)} = \frac{1}{\text{det}_{\mathbf{R}}(1-g\, q)}\,,
\end{eqnarray}
where we used the geometric series. Using the matrix identity $\log(\det(A)) = \tr(\log(A))$ and the logarithmic series $\log(1-x) = -\sum_{k=1}^\infty x^k / k$, we obtain the plethystic exponential
\begin{equation}
\sum_{d=0}^\infty \chi_{\text{Sym}^d(\mathbf{R})}(g) q^d = \text{exp}\bigg[\sum_{k=1}^\infty \frac{1}{k} q^k\, \text{Tr}_{\mathbf{R}}(g^k)\bigg]\,.
\end{equation}
%


\subsection{Fermionic Plethystic Exponential}

In the case of fermionic spurions we have to consider the antisymmetric tensor product, i.e.~we want to compute
\begin{equation}
\sum_{d=0}^\infty q^d\,  \chi_{\wedge^d\mathbf{R}} (g)\,,
\end{equation}
where $\wedge$ stands for the antisymmetric tensor product. We again pick the system of eigenvectors of $\mathbf{R}_V(g)$ as basis for $V$ and write $\mathbf{R}^{\otimes d}_{\wedge^dV}(g)\in GL(\wedge^dV)$ for the action of the group element $g$ on the vector space formed by the antisymmetric tensor product $\wedge^d V$. A basis for $\wedge^n V$ is $\lbrace \tfrac{1}{d!} \sum_{\sigma\in S_d} \epsilon(\sigma )\, e_{i_{\sigma(1)}}\otimes \ldots \otimes e_{i_{\sigma(d)}} | 1\leq i_1< \ldots < i_d\leq n\rbrace$, where $\epsilon(\sigma)$ returns the sign of the permutation. Coming back to the example with $d=2$ and $n=3$, the basis for $\wedge^2 V$ is
\begin{equation}
\lbrace e_1\wedge e_2 , e_1\wedge e_3 , e_2\wedge e_3\rbrace = \lbrace \tfrac{1}{2} (e_1\otimes e_2 - e_2\otimes e_1) , \tfrac{1}{2} (e_1\otimes e_3 - e_3\otimes e_1), \tfrac{1}{2} (e_2\otimes e_3 - e_3\otimes e_2)\rbrace\,,
\end{equation}
with the group character $\chi_{\wedge^2\mathbf{R}}(g)$ given by
\begin{equation}
\chi_{\wedge^2\mathbf{R}}(g) = \text{Tr}\big(\mathbf{R}^{\otimes 2}_{\wedge^2(V)}(g)\big) = \lambda_1\lambda_2 + \lambda_1\lambda_3 + \lambda_2\lambda_3 = \tfrac{1}{2} (\chi_{\mathbf{R}}(g)^2 - \chi_{\mathbf{R}}(g^2))\,.
\end{equation}
For general $d$ and $n$ each basis element of $\wedge^d V$ contains $d$ different basis elements $e_k$. This implies that if $d>n$ then $\wedge^d V$ is an empty space. The group character is the sum over the eigenvalues
\begin{equation}
\chi_{\wedge^d\mathbf{R}}(g) = \text{Tr}\big(\mathbf{R}^{\otimes d}_{\wedge^dV}(g)\big) = \sum_{1\leq i_1< \ldots < i_d\leq n} \lambda_{i_1}\cdots \lambda_{i_d}\,.
\end{equation}
Summing over $d$ we obtain the fermionic plethystic exponential
\begin{eqnarray}
\sum_{d=0}^\infty q^d\,  \chi_{\wedge^d(\mathbf{R})} (g) &=& \sum_{d=0}^\infty q^d \sum_{1\leq i_1< \ldots < i_d\leq n} \lambda_{i_1}\cdots \lambda_{i_d} = \prod_{i=1}^n (1+ \lambda_i\, q) = \text{det}(\mathbbm{1} + \mathbf{R}_V(g)\, q) \nonumber \\
&=& \text{det}_{\mathbf{R}}(1 + g\, q) = \text{exp}\bigg[\sum_{k=1}^\infty \frac{(-1)^{k+1}}{k} q^k\, \text{Tr}_{\mathbf{R}}(g^k)\bigg]\,,
\end{eqnarray}
where we again used the matrix identity $\log(\det(A)) = \tr(\log(A))$ and the logarithmic series $\log(1+x) = \sum_{k=1}^\infty (-1)^{k+1} x^k / k$ in the last line.
%


\section{Dimension 8 GRSMEFT basis}\label{app:GRSMEFT8}

We compile in Tables~\ref{tab:dim8Boson} and \ref{tab:dim8Ferm} the explicit operator basis for the GRSMEFT at mass dimension 8.

We recall that we constructed the Hilbert series in terms of the chiral components of the gauge field strengths ($X_{L/R}$) and the Weyl tensor ($C_{L/R}$), which are related to the standard field strengths and their duals by
\begin{equation}
X_{L/R}^{\mu\nu} = \frac{1}{2} \left( X^{\mu\nu} \pm i \tilde{X}^{\mu\nu}\right)\,,\quad C_{L/R}^{\mu\nu\rho\sigma} = \frac{1}{2}\left( C^{\mu\nu\rho\sigma} \pm i \tilde{C}^{\mu\nu\rho\sigma}\right)\,,
\end{equation}
with $\tilde{X}^{\mu\nu}= \tfrac{1}{2}\epsilon^{\mu\nu\alpha\beta} X_{\alpha\beta}$. For the dual of the Weyl tensor we can define a left- and a right-dual tensor $\,^*C^{\mu\nu\rho\sigma} = \tfrac{1}{2} \epsilon^{\mu\nu\alpha\beta} C_{\alpha\beta}\,^{\rho\sigma}$ and $C^{\mu\nu\rho\sigma}\,^* = \tfrac{1}{2} \epsilon^{\rho\sigma\alpha\beta} C^{\mu\nu}\,_{\alpha\beta}$, however one can show that $\,^*C^{\mu\nu\rho\sigma} = C^{\mu\nu\rho\sigma}\,^*$ and therefore we can define without ambiguity $\tilde{C}^{\mu\nu\rho\sigma} = \tfrac{1}{2} \epsilon^{\mu\nu\alpha\beta} C_{\alpha\beta}\,^{\rho\sigma}$.
Useful relations to trade chiral components for the standard field strength and its dual are
\begin{eqnarray}
C_{L/R\, \mu\nu\rho\sigma} C_{L/R}^{\mu\nu\rho\sigma} &=& \frac{1}{2} \left( C_{\mu\nu\rho\sigma} C^{\mu\nu\rho\sigma} \pm i C_{\mu\nu\rho\sigma} \tilde{C}^{\mu\nu\rho\sigma} \right)\,, \\
X_{L/R\, \rho\sigma} C_{L/R}^{\mu\nu\rho\sigma} &=& \frac{1}{2} \left( X_{\rho\sigma} C^{\mu\nu\rho\sigma} \pm i \tilde{X}_{\rho\sigma} C^{\mu\nu\rho\sigma}\right)\,.
\end{eqnarray}
We also use that $\epsilon_{i_1\ldots i_n} \epsilon^{j_1\ldots j_n} = n! \delta^{j_1}_{[ i_i} \cdots \delta^{j_n}_{ i_n ]}$ to rewrite pairs of building blocks that both include a dual field strength in terms of contractions without the Levi-Civita tensor.
\begin{table}[t!]
\centering
\begin{tabular}{||c|c||c|c||}
  \hline \hline
  \multicolumn{2}{||c||}{$C^4$} & \multicolumn{2}{c||}{$H^2 C^3$} \\ \cline{1-2} \cline{3-4}
  $\mathcal{O}_{CC}$ & $(C_{\mu\nu\rho\sigma} C^{\mu\nu\rho\sigma})^2$  & $\mathcal{O}_{2HC}$ & $(H^\dagger H) (C_{\mu\nu\rho\sigma} C^{\rho\sigma\alpha\beta} C_{\alpha\beta}\,^{\mu\nu})$  \\
   $\mathcal{O}_{C\tilde{C}}$ & $(C_{\mu\nu\rho\sigma} C^{\mu\nu\rho\sigma})(C_{\alpha\beta\gamma\delta} \tilde{C}^{\alpha\beta\gamma\delta})$  & $\mathcal{O}_{2H\tilde{C}}$ & $(H^\dagger H) (C_{\mu\nu\rho\sigma} C^{\rho\sigma\alpha\beta} \tilde{C}_{\alpha\beta}\,^{\mu\nu})$  \\
   $\mathcal{O}_{\tilde{C}\tilde{C}}$ & $(C_{\mu\nu\rho\sigma} \tilde{C}^{\mu\nu\rho\sigma})^2$  & ~ & ~ \\
  \hline \hline
  \multicolumn{2}{||c||}{$H^4 C^2$} & \multicolumn{2}{c||}{$X^3 C$} \\ \cline{1-2} \cline{3-4}
  $\mathcal{O}_{4HC}$  & $(H^\dagger H)^2 (C_{\mu\nu\rho\sigma} C^{\mu\nu\rho\sigma})$ & $\mathcal{O}_{GBC}$  & $G^A_{\mu\nu} G^A_{\rho\alpha} B^\alpha\,_\sigma C^{\mu\nu\rho\sigma}$\\
  $\mathcal{O}_{4H\tilde{C}}$  & $(H^\dagger H)^2 (C_{\mu\nu\rho\sigma} \tilde{C}^{\mu\nu\rho\sigma})$ & $\mathcal{O}_{GB\tilde{C}}$  & $G^A_{\mu\nu} G^A_{\rho\alpha} B^\alpha\,_\sigma \tilde{C}^{\mu\nu\rho\sigma}$\\
  ~  & ~ & $\mathcal{O}_{WBC}$  & $W^a_{\mu\nu} W^a_{\rho\alpha} B^\alpha\,_\sigma C^{\mu\nu\rho\sigma}$\\
  ~ & ~ & $\mathcal{O}_{WB\tilde{C}}$  & $W^a_{\mu\nu} W^a_{\rho\alpha} B^\alpha\,_\sigma \tilde{C}^{\mu\nu\rho\sigma}$\\
  \hline \hline
  \multicolumn{4}{||c||}{$X^2 C^2$}  \\ \cline{1-4} 
  $\mathcal{O}_{GC}^{(1)}$ & $(G^A_{\mu\nu} G^{A\, \mu\nu})(C_{\alpha\beta\rho\sigma}C^{\alpha\beta\rho\sigma})$  & $\mathcal{O}_{G\tilde{C}}^{(1)}$ & $(G^A_{\mu\nu} G^{A\, \mu\nu})(C_{\alpha\beta\rho\sigma}\tilde{C}^{\alpha\beta\rho\sigma})$ \\
  $\mathcal{O}_{\tilde{G}C}$ &  $(G^A_{\mu\nu} \tilde{G}^{A\, \mu\nu})(C_{\alpha\beta\rho\sigma}C^{\alpha\beta\rho\sigma})$ & $\mathcal{O}_{\tilde{G}\tilde{C}}$ &  $(G^A_{\mu\nu} \tilde{G}^{A\, \mu\nu})(C_{\alpha\beta\rho\sigma}\tilde{C}^{\alpha\beta\rho\sigma})$  \\
  $\mathcal{O}_{GC}^{(2)}$ & $G^A_{\mu\nu}G^{A\, \rho\sigma} C^{\mu\nu\alpha\beta}C_{\alpha\beta\rho\sigma}$  & $\mathcal{O}_{G\tilde{C}}^{(2)}$ & $G^A_{\mu\nu}G^{A\, \rho\sigma} C^{\mu\nu\alpha\beta}\tilde{C}_{\alpha\beta\rho\sigma}$   \\
$\mathcal{O}_{WC}^{(1)}$ & $(W^a_{\mu\nu} W^{a\, \mu\nu})(C_{\alpha\beta\rho\sigma}C^{\alpha\beta\rho\sigma})$  & $\mathcal{O}_{W\tilde{C}}^{(1)}$ & $(W^a_{\mu\nu} W^{a\, \mu\nu})(C_{\alpha\beta\rho\sigma}\tilde{C}^{\alpha\beta\rho\sigma})$  \\
  $\mathcal{O}_{\tilde{W}C}$ &  $(W^a_{\mu\nu} \tilde{W}^{a\, \mu\nu})(C_{\alpha\beta\rho\sigma}C^{\alpha\beta\rho\sigma})$ & $\mathcal{O}_{\tilde{W}\tilde{C}}$ &  $(W^a_{\mu\nu} \tilde{W}^{a\, \mu\nu})(C_{\alpha\beta\rho\sigma}\tilde{C}^{\alpha\beta\rho\sigma})$ \\
  $\mathcal{O}_{WC}^{(2)}$ & $W^a_{\mu\nu}W^{a\, \rho\sigma} C^{\mu\nu\alpha\beta}C_{\alpha\beta\rho\sigma}$  & $\mathcal{O}_{W\tilde{C}}^{(2)}$ & $W^a_{\mu\nu}W^{a\, \rho\sigma} C^{\mu\nu\alpha\beta}\tilde{C}_{\alpha\beta\rho\sigma}$   \\
  $\mathcal{O}_{BC}^{(1)}$ & $(B_{\mu\nu} B^{\mu\nu})(C_{\alpha\beta\rho\sigma}C^{\alpha\beta\rho\sigma})$  & $\mathcal{O}_{B\tilde{C}}^{(1)}$ & $(B_{\mu\nu} B^{\mu\nu})(C_{\alpha\beta\rho\sigma}\tilde{C}^{\alpha\beta\rho\sigma})$  \\
  $\mathcal{O}_{\tilde{B}C}$ &  $(B_{\mu\nu} \tilde{B}^{\mu\nu})(C_{\alpha\beta\rho\sigma}C^{\alpha\beta\rho\sigma})$ & $\mathcal{O}_{\tilde{B}\tilde{C}}$ &  $(B_{\mu\nu} \tilde{B}^{\mu\nu})(C_{\alpha\beta\rho\sigma}\tilde{C}^{\alpha\beta\rho\sigma})$  \\
  $\mathcal{O}_{BC}^{(2)}$ & $B_{\mu\nu}B^{\rho\sigma} C^{\mu\nu\alpha\beta}C_{\alpha\beta\rho\sigma}$  & $\mathcal{O}_{B\tilde{C}}^{(2)}$ & $B_{\mu\nu}B^{\rho\sigma} C^{\mu\nu\alpha\beta}\tilde{C}_{\alpha\beta\rho\sigma}$   \\
  \hline \hline
  \multicolumn{4}{||c||}{$X^2 H^2 C$}  \\ \cline{1-4}
  $\mathcal{O}_{GHC}$ & $(H^\dagger H) (G^{A\,\mu\nu} G^{A\, \rho\sigma} C_{\mu\nu\rho\sigma})$ & $\mathcal{O}_{GH\tilde{C}}$ & $(H^\dagger H) (G^{A\,\mu\nu} G^{A\, \rho\sigma} \tilde{C}_{\mu\nu\rho\sigma})$\\
  $\mathcal{O}_{WHC}$ & $(H^\dagger H) (W^{a\,\mu\nu} W^{a\, \rho\sigma} C_{\mu\nu\rho\sigma})$ & $\mathcal{O}_{WH\tilde{C}}$ & $(H^\dagger H) (W^{a\,\mu\nu} W^{a\, \rho\sigma} \tilde{C}_{\mu\nu\rho\sigma})$ \\
  $\mathcal{O}_{BHC}$ & $(H^\dagger H) (B^{\mu\nu} B^{\rho\sigma} C_{\mu\nu\rho\sigma})$ & $\mathcal{O}_{BH\tilde{C}}$ & $(H^\dagger H) (B^{\mu\nu} B^{\rho\sigma} \tilde{C}_{\mu\nu\rho\sigma})$\\
  $\mathcal{O}_{WBC}$ & $(H^\dagger \tau^a H) (B^{\mu\nu} W^{a\, \rho\sigma} C_{\mu\nu\rho\sigma})$ & $\mathcal{O}_{WB\tilde{C}}$ & $(H^\dagger \tau^a H) (B^{\mu\nu} W^{a\, \rho\sigma} \tilde{C}_{\mu\nu\rho\sigma})$ \\
  \hline\hline
  \multicolumn{2}{||c||}{$H^2 C^2 \mathcal{D}^2$} & \multicolumn{2}{c||}{$X H^2 C \mathcal{D}^2$} \\ \cline{1-2} \cline{3-4}
     $\mathcal{O}_{HC\mathcal{D}}$ & $(\nabla_\alpha H)^\dagger (\nabla^\alpha H) (C_{\mu\nu\rho\sigma} C^{\mu\nu\rho\sigma})$ & $\mathcal{O}_{WC\mathcal{D}}$  & $(\nabla^\mu H)^\dagger \tau^a (\nabla^\nu H) W^{a\, \rho\sigma} C_{\rho\sigma\mu\nu}$ \\
      $\mathcal{O}_{H\tilde{C}\mathcal{D}}$ & $(\nabla_\alpha H)^\dagger (\nabla^\alpha H) (C_{\mu\nu\rho\sigma} \tilde{C}^{\mu\nu\rho\sigma})$ & $\mathcal{O}_{W\tilde{C}\mathcal{D}}$  & $(\nabla^\mu H)^\dagger \tau^a (\nabla^\nu H) W^{a\, \rho\sigma} \tilde{C}_{\rho\sigma\mu\nu}$ \\
     ~ & ~ & $\mathcal{O}_{BC\mathcal{D}}$  & $(\nabla^\mu H)^\dagger (\nabla^\nu H) B^{\rho\sigma} C_{\rho\sigma\mu\nu}$ \\
     ~ & ~ & $\mathcal{O}_{B\tilde{C}\mathcal{D}}$  & $(\nabla^\mu H)^\dagger (\nabla^\nu H) B^{\rho\sigma} \tilde{C}_{\rho\sigma\mu\nu}$ \\
  \hline \hline
 \end{tabular}
\caption{Bosonic dimension-8 operators of the GRSMEFT including gravitational interactions.}
\label{tab:dim8Boson}
\end{table}
\begin{table}[t!]
\centering
\begin{tabular}{||c|c||c|c||}
  \hline \hline
  \multicolumn{2}{||c||}{$\psi^2 H C^2$} & \multicolumn{2}{c||}{$\psi^2 XHC$}  \\ \cline{1-2} \cline{3-4}
  $\mathcal{O}_{uHC}$	& 	$(\bar{Q}_L \tilde{H} u_R) (C_{\mu\nu\rho\sigma} C^{\mu\nu\rho\sigma})$ 		& 	$\mathcal{O}_{uGC}$ 	& $(\bar{Q}_L \sigma^{\mu\nu} T^A u_R) \tilde{H} G^{A\, \rho\sigma}C_{\mu\nu\rho\sigma}$\\
  $\mathcal{O}_{uH\tilde{C}}$	& 	$(\bar{Q}_L \tilde{H} u_R) (C_{\mu\nu\rho\sigma} \tilde{C}^{\mu\nu\rho\sigma})$ 		& $\mathcal{O}_{uWC}$ & $(\bar{Q}_L \sigma^{\mu\nu} u_R) \tau^a \tilde{H} W^{a\, \rho\sigma}C_{\mu\nu\rho\sigma}$ \\
  $\mathcal{O}_{dHC}$	& 	$(\bar{Q}_L H d_R) (C_{\mu\nu\rho\sigma} C^{\mu\nu\rho\sigma})$ 		& $\mathcal{O}_{uBC}$ & $(\bar{Q}_L \sigma^{\mu\nu} u_R) \tilde{H} B^{\rho\sigma} C_{\mu\nu\rho\sigma}$   \\
  $\mathcal{O}_{dH\tilde{C}}$	& 	$(\bar{Q}_L H d_R) (C_{\mu\nu\rho\sigma} \tilde{C}^{\mu\nu\rho\sigma})$ 		& $\mathcal{O}_{dGC}$ & $(\bar{Q}_L \sigma^{\mu\nu} T^A d_R) H G^{A\, \rho\sigma} C_{\mu\nu\rho\sigma}$  \\
  $\mathcal{O}_{eHC}$	& 	$(\bar{L}_L H e_R) (C_{\mu\nu\rho\sigma} C^{\mu\nu\rho\sigma})$ 		& $\mathcal{O}_{dWC}$ & $(\bar{Q}_L \sigma^{\mu\nu} d_R)\tau^a H W^{a\, \rho\sigma}C_{\mu\nu\rho\sigma}$  \\
  $\mathcal{O}_{eH\tilde{C}}$	& 	$(\bar{L}_L H e_R) (C_{\mu\nu\rho\sigma} \tilde{C}^{\mu\nu\rho\sigma})$ 		& $\mathcal{O}_{dBC}$ & $(\bar{Q}_L \sigma^{\mu\nu} d_R) H B^{\rho\sigma} C_{\mu\nu\rho\sigma}$  \\
    ~	& 	~		& $\mathcal{O}_{eWC}$ & $(\bar{L}_L \sigma^{\mu\nu} e_R) \tau^a H W^{a\, \rho\sigma} C_{\mu\nu\rho\sigma}$  \\
  ~	& 	~ 		& $\mathcal{O}_{eBC}$ & $(\bar{L}_L \sigma^{\mu\nu} e_R) H B^{\rho\sigma} C_{\mu\nu\rho\sigma}$  \\
  \hline \hline
  \multicolumn{2}{||c||}{$\psi^2 H C \mathcal{D}^2$} & \multicolumn{2}{c||}{$\psi^4 C$}  \\ \cline{1-2} \cline{3-4}
  $\mathcal{O}_{uC\mathcal{D}}$  & $(\bar{Q}_L \sigma^{\mu\nu} \nabla^\rho u_R) (\nabla^\sigma \tilde{H}) C_{\mu\nu\rho\sigma}$ 	& $\mathcal{O}_{udC}$	& $\epsilon_{ij} (\bar{Q}_L^i \sigma^{\mu\nu} u_R) (\bar{Q}_L^j \sigma^{\rho\sigma} d_R) C_{\mu\nu\rho\sigma}$\\
  $\mathcal{O}_{dC\mathcal{D}}$  & $(\bar{Q}_L \sigma^{\mu\nu} \nabla^\rho d_R) (\nabla^\sigma H) C_{\mu\nu\rho\sigma}$	& $\mathcal{O}_{ueC}$	& $\epsilon_{ij}(\bar{Q}_L^i \sigma^{\mu\nu}u_R) (\bar{L}_L^j \sigma^{\rho\sigma} e_R) C_{\mu\nu\rho\sigma}$	\\
  $\mathcal{O}_{eC\mathcal{D}}$	&	$(\bar{L}_L \sigma^{\mu\nu} \nabla^\rho e_R) (\nabla^\sigma H) C_{\mu\nu\rho\sigma}$	& $\mathcal{O}_{uedC}$	& $\epsilon^{\alpha\beta\gamma} [(d_R^\alpha)^T C \sigma^{\mu\nu} u_R^\beta ] [(u_R^\gamma)^T C \sigma^{\rho\sigma} e_R] C_{\mu\nu\rho\sigma}$\\
    \hline \hline
  \multicolumn{4}{||c||}{$\psi^2 X C \mathcal{D}$}   \\ \cline{1-4}
  $\mathcal{O}_{QGC\mathcal{D}}$	& $(\bar{Q}_L \gamma^\mu T^A \nabla^\nu Q_L) G^{A\, \rho\sigma} C_{\mu\nu\rho\sigma}$	&	$\mathcal{O}_{QG\tilde{C}\mathcal{D}}$	& $(\bar{Q}_L \gamma^\mu T^A \nabla^\nu Q_L) G^{A\, \rho\sigma} \tilde{C}_{\mu\nu\rho\sigma}$	\\
  $\mathcal{O}_{uGC\mathcal{D}}$	&	$(\bar{u}_R \gamma^\mu T^A \nabla^\nu u_R) G^{A\, \rho\sigma} C_{\mu\nu\rho\sigma}$	&	$\mathcal{O}_{uG\tilde{C}\mathcal{D}}$	& $(\bar{u}_R \gamma^\mu T^A \nabla^\nu u_R) G^{A\, \rho\sigma} \tilde{C}_{\mu\nu\rho\sigma}$	\\
   $\mathcal{O}_{dGC\mathcal{D}}$	&	$(\bar{d}_R \gamma^\mu T^A \nabla^\nu d_R) G^{A\, \rho\sigma} C_{\mu\nu\rho\sigma}$	&	$\mathcal{O}_{dG\tilde{C}\mathcal{D}}$	& $(\bar{d}_R \gamma^\mu T^A \nabla^\nu d_R) G^{A\, \rho\sigma} \tilde{C}_{\mu\nu\rho\sigma}$	\\
   $\mathcal{O}_{QWC\mathcal{D}}$	&	$(\bar{Q}_L \gamma^\mu \tau^a \nabla^\nu Q_L) W^{a\, \rho\sigma} C_{\mu\nu\rho\sigma}$ & $\mathcal{O}_{QW\tilde{C}\mathcal{D}}$	&	$(\bar{Q}_L \gamma^\mu \tau^a \nabla^\nu Q_L) W^{a\, \rho\sigma} \tilde{C}_{\mu\nu\rho\sigma}$\\
   $\mathcal{O}_{LWC\mathcal{D}}$	&	$(\bar{L}_L \gamma^\mu \tau^a \nabla^\nu L_L) W^{a\, \rho\sigma} C_{\mu\nu\rho\sigma}$ & $\mathcal{O}_{LW\tilde{C}\mathcal{D}}$	&	$(\bar{L}_L \gamma^\mu \tau^a \nabla^\nu L_L) W^{a\, \rho\sigma} \tilde{C}_{\mu\nu\rho\sigma}$\\
   $\mathcal{O}_{QBC\mathcal{D}}$	&	$(\bar{Q}_L \gamma^\mu \nabla^\nu Q_L) B^{\rho\sigma} C_{\mu\nu\rho\sigma}$ & $\mathcal{O}_{QB\tilde{C}\mathcal{D}}$	&	$(\bar{Q}_L \gamma^\mu \nabla^\nu Q_L) B^{\rho\sigma} \tilde{C}_{\mu\nu\rho\sigma}$\\
   $\mathcal{O}_{uBC\mathcal{D}}$	&	$(\bar{u}_R \gamma^\mu \nabla^\nu u_R) B^{\rho\sigma} C_{\mu\nu\rho\sigma}$ & $\mathcal{O}_{uB\tilde{C}\mathcal{D}}$	&	$(\bar{u}_R \gamma^\mu \nabla^\nu u_R) B^{\rho\sigma} \tilde{C}_{\mu\nu\rho\sigma}$\\
   $\mathcal{O}_{dBC\mathcal{D}}$	&	$(\bar{d}_R \gamma^\mu \nabla^\nu d_R) B^{\rho\sigma} C_{\mu\nu\rho\sigma}$ & $\mathcal{O}_{dB\tilde{C}\mathcal{D}}$	&	$(\bar{d}_R \gamma^\mu \nabla^\nu d_R) B^{\rho\sigma} \tilde{C}_{\mu\nu\rho\sigma}$\\
   $\mathcal{O}_{LBC\mathcal{D}}$	&	$(\bar{L}_L \gamma^\mu \nabla^\nu L_L) B^{\rho\sigma} C_{\mu\nu\rho\sigma}$ & $\mathcal{O}_{LB\tilde{C}\mathcal{D}}$	&	$(\bar{L}_L \gamma^\mu \nabla^\nu L_L) B^{\rho\sigma} \tilde{C}_{\mu\nu\rho\sigma}$\\
   $\mathcal{O}_{eBC\mathcal{D}}$	&	$(\bar{e}_R \gamma^\mu \nabla^\nu e_R) B^{\rho\sigma} C_{\mu\nu\rho\sigma}$ & $\mathcal{O}_{eB\tilde{C}\mathcal{D}}$	&	$(\bar{e}_R \gamma^\mu \nabla^\nu e_R) B^{\rho\sigma} \tilde{C}_{\mu\nu\rho\sigma}$\\
  \hline\hline
 \end{tabular}
\caption{Dimension-8 operators of the GRSMEFT including gravitational interactions and fermions for $N_f = 1$. We do not show explicitly the h.c.~of non self-conjugate operators. $\tilde{H}_i = \epsilon_{ij} H^*_j$ and $C$ is the charge conjugation matrix.}
\label{tab:dim8Ferm}
\end{table}


\section{Gravity EFT in $d>4$ Spacetime Dimensions%
\footnote{We thank B.~Henning for suggesting the study of this generalization to us.}}\label{app:HigherDim}
In Section~\ref{subsec:GravityBuildingBlocks} we identified the independent building blocks for the EFT of gravity, i.e.~the graviton single particle module $R_C$, in $d=4$ spacetime dimensions. However, this result does not only hold in 4 dimensions, but trivially extends to $d>4$.%
\footnote{In $d<4$ the Weyl tensor identically vanishes, i.e.~the graviton is not dynamical.} 
The derivation of the graviton single particle module only made use of the Einstein equations and Bianchi identities, which have the same form in any dimension. Consequently, also $R_C$ in Eq.~(\ref{eq:WeylSPM}) is valid in dimensions larger than 4.

In this appendix we will shortly outline how to obtain the character for the single particle module $\chi_C$ and the corresponding Hilbert series for gravity in vacuum in $d > 4$ and explicitly perform the construction for $d=5$.

\subsection{Character for the Single Particle Module}
We work in Euclidean space, where the Lorentz group in $d$ dimensions is $SO(d)$. The finite dimensional representations of $SO(d)$ can be labeled by partitions $l=(l_1, l_2,\ldots , l_r)$ with $l_1 \geq \cdots \geq l_{r-1} \geq |l_r|$ for $SO(2r)$ and $l_1 \geq \cdots \geq l_r \geq 0$ for $SO(2r+1)$, where $r$ is the rank of the group. These representations are in a one-to-one correspondence with Young diagrams, i.e.~they correspond to tensors with $|l| = \sum_i l_i$ indices, which are (anti)symmetrized according to the corresponding Young diagram (in this regard, such a labelling is more convenient than the one we have used for $SO(4)$ in the main text). In this notation, the fundamental representation is labeled by $l=(1,0,\ldots,0)$, the antisymmetric tensor with two indices is $l=(1,1,0,\ldots,0)$ and the completely symmetric, traceless tensor with $n$ indices corresponds to $l=(n,0,\ldots,0)$. Therefore the Weyl tensor in $d$ dimensions lives in the representation labeled by $l=(2,2,0,\ldots,0)$.%
\footnote{Note that $SO(2r)$ admits chiral representations if $l_r\neq 0$, which is why in $d=4$ the Weyl tensor can be decomposed into a left-handed and a right-handed part, which belong to the representations labeled by $(2,2)$ and $(2,-2)$, respectively.} 

The single particle module in Eq.~(\ref{eq:WeylSPM}) consists of the Weyl tensor and symmetric, traceless combinations of covariant derivatives acting on the Weyl tensor. In the notation we introduced above, it is clear that $\nabla_{\lbrace \mu_1}\cdots \nabla_{\mu_n} C_{\mu\rbrace\nu \rho\sigma}$ transforms in the representation corresponding to $l=(n+2,2,0,\ldots,0)$, which implies that its character is given by
\begin{equation}
\chi_C^{(d)} (\mathcal{D}; x) = \sum_{n=0}^\infty \mathcal{D}^{n+2}\chi^{(d)}_{(n+2,2,0,\ldots,0)} (x)\,,
\label{eq:charCd}
\end{equation}
where $\chi_l^{(d)} (x)$ are the $SO(d)$ characters.

Let us specialize to $d=5$ dimensions. Eq.~(\ref{eq:charCd}) can be evaluated explicitly using the Weyl character formula for $SO(5)$ characters, which can be found e.g.~in Appendix A of \cite{Henning:2017fpj}
\begin{equation}
\begin{split}
\chi_C^{(5)}(\mathcal{D};x) = \mathcal{D}^2 P^{(5)}(\mathcal{D};x)&\left[\chi^{(5)}_{(2,2)}(x) - \mathcal{D} \big(\chi^{(5)}_{(2,2)}(x) + \chi^{(5)}_{(2,1)}(x)\big)\right.\\
&\left.\quad+\mathcal{D}^2\big( \chi^{(5)}_{(2,1)}(x) + \chi^{(5)}_{(1,1)}(x) \big) - \mathcal{D}^3 \chi^{(5)}_{(1,1)}(x)\right]\,.
\end{split}
\label{eq:charC5}
\end{equation}
At this point we want to emphasize that unlike in $d=4$, the single particle module $R_C$ cannot be identified with a short conformal representation, whose scaling dimension saturates the unitarity bound, by formally assigning this scaling dimension to the Weyl tensor. 
The character for the conformal representation $[4;(2,2)]$ in $d=5$ dimensions is given by (see e.g.~\cite{Dolan:2005wy})
\begin{equation}
\chi_{[4;(2,2)]}^{(5)}(\mathcal{D};x) = \mathcal{D}^4 P^{(5)}(\mathcal{D};x)\left[ \chi_{(2,2)}^{(5)}(x) - \mathcal{D}\chi_{(2,1)}^{(5)}(x) + \mathcal{D}^2 \chi_{(1,1)}^{(5)}(x)\right]\,,
\label{eq:charConf5}
\end{equation}
which clearly differs from Eq.~(\ref{eq:charC5}) even after assigning a scaling dimension of $4$ to the Weyl tensor. The reason for this is that the conformal primary, i.e.~the equivalent of the Weyl tensor, does not satisfy the second Bianchi identity. The corresponding descendent $\nabla_{[\mu_1}C_{\mu\nu ]\rho\sigma}$, which transforms in the representation labeled by $l=(2,2)$, is therefore not subtracted from the conformal multiplet, as can be explicitly seen in Eq.~(\ref{eq:charConf5}).

Note that there is no known expression for $\Delta\mathcal{H}$ in Eq.~(\ref{eq:fullHS}) if the single particle modules cannot be identified with conformal representations. However, Eq.~(\ref{eq:HS}) for $\mathcal{H}_0$ still holds.

\subsection{Gravity in Vacuum in $d=5$}
The Hilbert series for pure gravity in $d=5$ dimensions can be computed using Eq.~(\ref{eq:HS}) and the character for the single particle module in Eq.~(\ref{eq:charC5}). Grading the spurions according to their mass dimension, i.e.~$C\rightarrow \epsilon^2 C$ and $\mathcal{D}\rightarrow \epsilon \mathcal{D}$, and expanding up to mass dimension ten we obtain\footnote{The explicit expressions for the $SO(5)$ integration measure and characters can be found e.g.~in the appendix of \cite{Henning:2017fpj}.}
\begin{equation}
\begin{split}
\mathcal{H}_0(\mathcal{D},C; \epsilon) &= \int d\mu_{SO(5)}(x) \frac{1}{P^{(5)}(\mathcal{\epsilon\, D},x)} \text{PE}\bigg[\frac{C}{\mathcal{D}^{2}}\bigg]\\
&=\epsilon^6 C^3 + \epsilon^8 4\,C^4 + \epsilon^{10}\left( 7\, C^4 \mathcal{D}^2 + 5\, C^5 \right) +\ldots\,,
\end{split}
\end{equation}
where we have dropped terms of mass dimension five or lower, since they also receive contributions from $\Delta\mathcal{H}$. Note that the number of independent operators in $d=5$ differs from the number of CP even operators in $d=4$, i.e.~the operators which can be written without the dual of the Weyl tensor. This implies that, beyond chirality, there are genuinely new contractions of Weyl tensors which are linearly dependent or vanish in $d=4$. The number of independent operators without derivatives can be cross-checked with the number of Weyl invariants given in \cite{Fulling:1992vm} and we find full agreement.

\clearpage



\end{document}